\documentclass [aps,prb,twocolumn,reprint,groupeaddress,showpacs]{revtex4-1}
\usepackage{graphicx}
\usepackage{gensymb}
\usepackage{amsmath}
\usepackage{bm}
	
\begin{document}
\preprint{Submitted to PRB} \title{Hyperuniform disordered phononic
  structures} \author{G.~Gkantzounis} \author{T.~Amoah}
\author{M.~Florescu} \affiliation{Department of Physics, Advanced
  Technology Institute, University of Surrey, Guildford, Surrey GU2
  7XH, UK} \date{\today}
\begin{abstract}
  We demonstrate the existence of large phononic band gaps in designed
  hyperuniform (isotropic) disordered two-dimensional (2D) phononic
  structures of Pb cylinders in epoxy matrix. The phononic band gaps
  in hyperuniform disordered phononic structures are comparable to
  band gaps of similar periodic structures, for both out-of-plane and
  in-plane polarizations. A large number of localized modes is
  identified near the band edges, as well as, diffusive transmission
  throughout the rest of the frequency spectrum. Very high-$Q$ cavity
  modes for both out-of-plane and in-plane polarizations are formed by
  selectively removing a single cylinder out of the
  structure. Efficient waveguiding with almost 100\% transmission
  trough waveguide structures with arbitrary bends is also
  presented. We expand our results to thin three-dimensional layers of
  such structures and demonstrate effective band gaps related to the
  respective 2D band gaps. Moreover, the drop in the $Q$ factor for
  the three-dimensional structures is not more than three orders of
  magnitude compared to the 2D ones.
\end{abstract}
\maketitle

\section{\label{sec:intro}Introduction}
Phononic crystals, artificial materials with periodically arranged
compounds, were introduced more than two decades ago as the elastic
waves analogue of photonic
crystals~\cite{sigalas_band_1993,kushwaha_bandgap_1994}. These
materials, either in two or three dimensions, are capable of exhibiting large
frequency regions of prohibited propagation of elastic waves, the
so-called phononic band gaps (PBGs). Phononic crystals have been
efficiently used in applications including audible
filters~\cite{vasseur_phononic_2002}, acoustic
diodes~\cite{li_tunable_2011} and cloaking~\cite{chen_acoustic_2010},
ultrasound imaging~\cite{sukhovich_negative_2008},
optomechanics~\cite{aspelmeyer_cavity_2014}, heat
conduction~\cite{maldovan_sound_2013,davis_nanophononic_2014}.

On the other hand, the hyperuniformity concept was first introduced as an order
metric for ranking point patterns according to their local density
fluctuations~\cite{torquato_local_2003}. Hyperuniform structures cover
the intermediate regime between random and periodic structures, and
exhibit properties usually associated with both of these two
extremes. Hyperuniform stealthy disordered photonic structures exhibit
large isotropic photonic band gaps and are capable of blocking light
of all
polarizations~\cite{florescu_designer_2009,man_isotropic_2013,muller_photonic_2014}.
Therefore, they can be efficiently used in the design of low
dielectric contrast band gaps \cite{man_photonic_2013}, high-$Q$
optical cavities \cite{florescu_optical_2013,amoah_high-$q$_2015},
free-form waveguides \cite{florescu_optical_2013,man_photonic_2013,Florescu2010},
polarizers \cite{zhou_hyperuniform_2016}, plasmon-enhanced Raman
spectroscopy \cite{zito_plasmon-enhanced_2014}, quantum cascade lasers
\cite{deglinnocenti_hyperuniform_2016}, etc.

Markedly, compared with the studies of disorder in photonic systems,
research on disordered phononic systems has been sparse. There have
been some studies focused on two-dimensional
(2D)~\cite{zuo-dong_elastic_2005,wagner_2d_2015} and three-dimensional
(3D)~\cite{limonov_optical_2012,sainidou_widening_2005,still_simultaneous_2008}
disordered phononic structures, but these structures were not ideally
suited for opening large phononic band gaps (PBGs). The most
insightful study into disorder has been by Chen et
al.~\cite{chen_study_2007}. The study of disorder in elastic media is
relevant for macroscopic scale applications, such as control of
seismic waves~\cite{larose_weak_2004} and modeling of
rocks~\cite{shahbazi_localization_2005} to the micro- and nano-scale
applications, including acoustic filters~\cite{tourin_time_2006-1},
piezoelectric
materials~\cite{li_frequency-dependent_2007,wang_propagation_2008},
aviation~\cite{lin_shape_????},
bio-materials~\cite{davies_hypothesis:_2014},
fracture~\cite{davies_hypothesis:_2014,shekhawat_damage_2013},
manipulation of the thermal
conductance~\cite{zen_engineering_2014,maldovan_sound_2013,oconnor_heat_1974,maire_thermal_2015},
etc. Moreover appropriately designed disordered phononic systems can
suppress wave transport for a wider range of frequencies than their
periodic counterparts~\cite{sainidou_widening_2005}. Designed
hyperuniform disordered phononic structures (HDPS) have the advantage
of exploiting both structural disorder, due to the absence of
translation symmetry, and sort-scale ordering, due to the high degree
of hyperuniformity. This combination makes them ideal for phonon
manipulation since they can exhibit large band gaps, isotropy and
diffusive propagation at the same time. Therefore they can play a very
important role in all the above mentioned application areas.

In this paper we introduce and analyze HDPS. Such structures exhibit
structural disorder, while being statistically
isotropic~\cite{torquato_local_2003,uche_constraints_2004}. Moreover,
they are designed from totally disordered initial states, therefore
they have no sign of underlying periodicity, which differentiates them
from all previous structures studied in the literature. The radial
distribution function and the diffraction patterns (2D Fourier
transform) of the structure are used to investigate the correlated
disorder of the structure. The material parameters used are the same
as those used by Chen at
al. \cite{chen_study_2007,chen_localisation_2010}. In
Sec.~\ref{sec:hpuPnBG} the band structure of such HPDS is discussed
for all polarizations. Large phononic band gaps, similar to the
periodic case, are observed. To identify the origin of these gaps we
introduce the concentration factor, previously used in the photonic
hyperuniform structures, for both out-of-plane (pure transverse) and
in-plane (mixed longitudinal-transverse) elastic modes. Localized
modes and propagation of elastic waves in these structures are
thoroughly investigated. In Sec.~\ref{sec:CavWG} high-$Q$ cavity modes
are introduced into the structures by removing single cylinders and
thoroughly analyzed. Moreover, waveguiding through arbitrary shaped
waveguides with HDPS-like walls is extensively
discussed. Sec.~\ref{sec:3D} extends the previous results in finite
thickness 3D slabs.
 
\section{\label{sec:hpuPnBG}Frequency Bands of 2D structures}

A hyperuniform point pattern is a  point pattern in real space
for which the number variance $\sigma^2(R)$ within a spherical
sampling window of radius $R$ (in $d$ dimensions) grows more slowly
than the window volume ($\propto R^d$) for large $R$. We 
furthermore consider that the point pattern is stealthy, i.e., the
structure factor $S(\mathbf{k})$, defined
as~\cite{torquato_local_2003}
\begin{equation}\label{eq:SF}
  S(\mathbf{k}) = \frac{1}{N}\left|\sum_{n=1}^{N} 
    e^{\text{i}\mathbf{k}\cdot\mathbf{r}_n}\right|^2\;,
\end{equation}
where $\mathbf{k}$ are vectors in the reciprocal space and
$\mathbf{r}_n$, $n=1,\ldots,N$ are the positions of the $N$ particles,
is isotropic and vanishes for a finite range of wave numbers $0<k\leq
k_0$ for some positive critical wave vector,
$k_0$~\cite{batten_classical_2008}. The size of this region can be
expressed through the so-called stealthy parameter
$\chi=M(\mathbf{k})/dN$, where $M(\mathbf{k})$ the number of linearly
independent $k$ vectors where $S(\mathbf{k})=0$ and $d=2$ in the
present case~\cite{batten_classical_2008,florescu_designer_2009}.

\begin{figure}[h]
  \centerline{\includegraphics*[width=\linewidth]{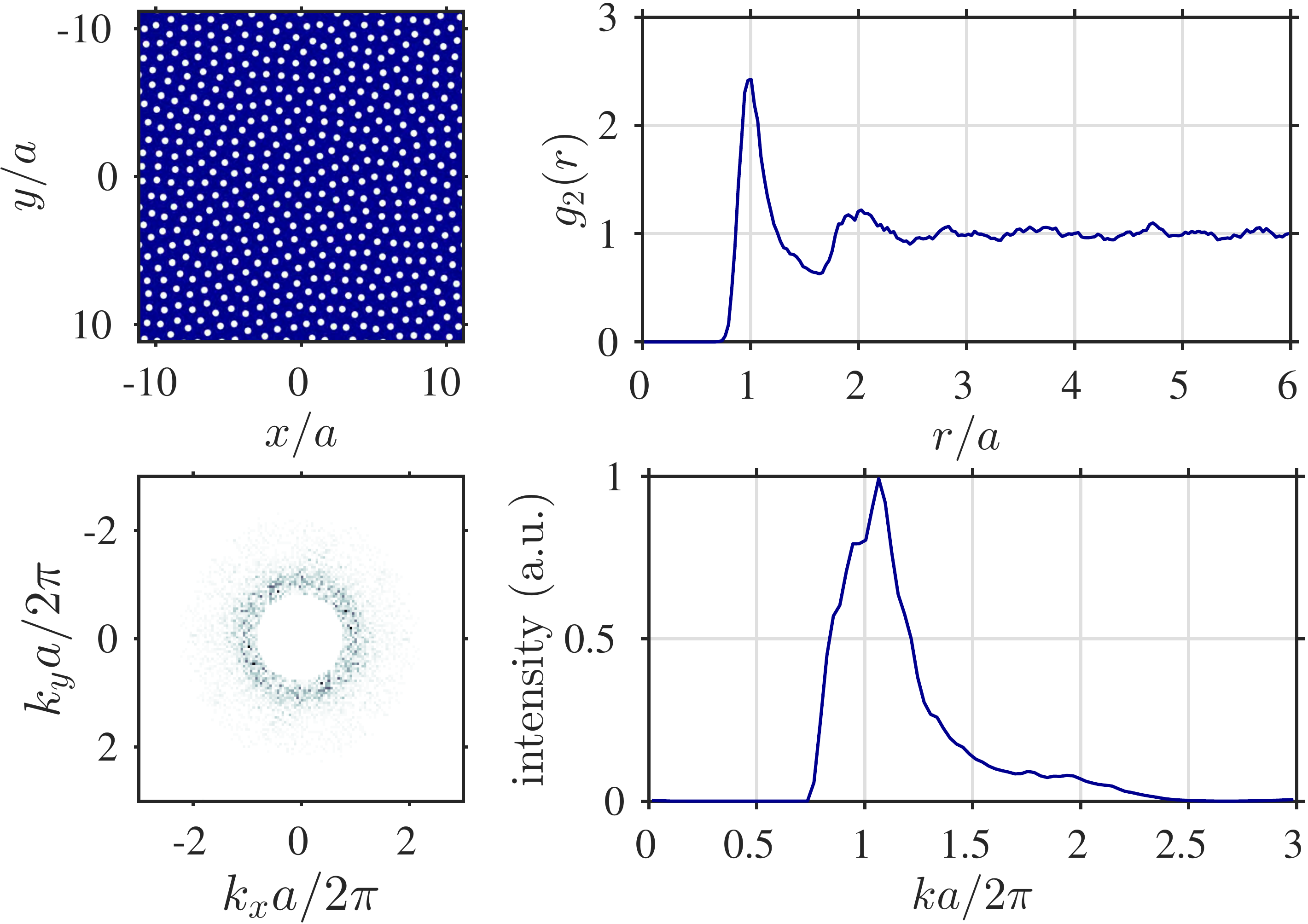}}
  \caption{\label{fig:SuperCell} (Color online) \textbf{(a)} A super
    cell of the proposed HDPS, consisting of 500 cylinders (radius
    0.25$a$, $\chi=0.5$). \textbf{(b)} The radial distribution
    function, $g_2(r)$ of the hyperuniform points
    pattern. \textbf{(c)} The magnitude of the Fourier transform
    components of the HDPS. \textbf{(d)} The radially integrated
    Fourier components. We can clearly identify the region where these
    components and the corresponding structure factor vanish, below
    $k_0=0.71(2\pi/a)$ }
\end{figure}
We consider, as a working example, an HDPS consisting of $N=500$
identical cylinders with a radius $0.25a$, $a$ being the average
distance among the centers of the cylinders, distributed according to
a stealthy hyperuniform point pattern, as shown in
Fig.~\ref{fig:SuperCell}(a), with $\chi=0.5$. We consider a supercell
of the structure with dimensions $\sqrt{N} a \times \sqrt{N}
a$. Fig.~\ref{fig:SuperCell}(b) depicts the radial distribution
function, $g_2(r)$~\cite{torquato_local_2003}. We can clearly identify
two peaks corresponding to first and second neighbor average
distance. Fig.~\ref{fig:SuperCell}(c) depicts the magnitude of the 2D
Fourier components (structure factor) of the image of structure
(Fig.~\ref{fig:SuperCell}(a)), i.e. the structure factor of the
decorated point pattern. This structure factor differs from the
structure factor of the point pattern only at the high-$k$ values,
i.e. outside the region of the ring we observe in
Fig.~\ref{fig:SuperCell}(c). The ``stealthiness'' of the structure can
be clearly identified by the radial symmetry and the region of
vanishing structure factor within a circle of radius $k_0$ with
$k_0a/2\pi=0.71$, as shown in Fig.~\ref{fig:SuperCell}(d).

Let us now consider the propagation of elastic waves through the
structure. We are interested in the frequency-domain response of the
structures, i.e., we solve for fields of the form
$\mathbf{u}(\mathbf{r},t)=\Re[\mathbf{u}(\mathbf{r})\exp(-\mathrm{i}\omega
t)]$, where $\mathbf{u}(\mathbf{r})$ the (complex) time-independent
elastic field component at position $\mathbf{r}$ and $\omega=2\pi f$
the angular frequency. The time-independent wave equation in an
inhomogeneous isotropic medium characterized by position-dependent
mass density $\rho$ and position-dependent Lam\`{e} coefficients
$\lambda$ and $\mu$ takes the form~\cite{chen_localisation_2010}
\begin{equation}\label{eq:Elastic3D}
  \nabla \left( \lambda \nabla \cdot \mathbf{u} \right) + \nabla 
  \cdot \left[ \mu \left( \nabla\mathbf{u} +  \nabla
      \mathbf{u}^{\mathrm{T}} \right) \right] = -\rho \omega^2 
  \mathbf{u}
\end{equation}
where $\nabla\mathbf{u}$ the tensor gradient of the displacement field
and $\nabla\mathbf{u}^{\mathrm{T}}$ the transpose of the tensor
gradient. A 2D structure is composed of infinite-height cylinders,
i.e. the system is considered to be homogeneous along the $z$
direction and propagation is restricted in the $xy$ plane. In this
case Eq.~(\ref{eq:Elastic3D}) splits into two independent sets of
equations, namely,
\begin{equation}\label{eq:ElasticZ}
  \nabla \cdot [ \mu \nabla u_z ] = - \rho \omega^2 u_z 
  \;,
\end{equation}
for elastic waves polarized along the $z$ axis and
\begin{eqnarray}\label{eq:ElasticXY}
  \nabla_{\parallel} \left( \lambda \nabla_{\parallel} \cdot 
    \mathbf{u}_\parallel \right) &+&
  \nabla_\parallel \cdot \left[ \mu 
    \left( \nabla_\parallel\mathbf{u}_\parallel + \nabla_\parallel 
      \mathbf{u}_\parallel^{\mathrm{T}} \right) 
  \right] \nonumber\\
  &=& -\rho \omega^2 \mathbf{u}_\parallel
\end{eqnarray}
for elastic waves polarized parallel to the $xy$ plane, where
$\mathbf{u}_\parallel = \hat{\mathbf{x}} u_x + \hat{\mathbf{y}} u_y $
and $ \nabla_\parallel = \hat{\mathbf{x}} (\partial/\partial x) +
\hat{\mathbf{y}} (\partial/\partial y) $. We note that
Eq.~(\ref{eq:ElasticZ}) is a pseudo-scalar equation for the $u_z$
component of the purely transverse elastic field, however
Eq.~(\ref{eq:ElasticXY}) couples the $u_x$ and $u_y$ components of the
elastic field. Moreover, $\mathbf{u}_{\parallel}$ is a mixture of a
longitudinal and a transverse component. In order to separate the
longitudinal and transverse components we can use the displacement
potentials $\phi(\mathbf{r})$ and
$\mathbf{A}(\mathbf{r})$, which are defined by the equation
\begin{equation}\label{eq:Potentials}
  \mathbf{u} (\mathbf{r}) = \nabla \phi(\mathbf{r}) + \nabla \times 
  \mathbf{A} (\mathbf{r})\;.
\end{equation}
The wave equations satisfied by the displacement potentials,
$\phi(\mathbf{r})$ and $\mathbf{A}(\mathbf{r})$, can be written in the
form
\begin{equation}\label{eq:WavePhi}
  \nabla^{2} \phi (\mathbf{r}) + \frac{\omega^2}{c_l^2} \phi (\mathbf{r})
  =0\;,
\end{equation}
and
\begin{equation}\label{eq:WaveAlpha}
  \nabla^{2} \mathbf{A} (\mathbf{r}) + 
  \frac{\omega^2}{c_t^2} \mathbf{A} (\mathbf{r}) = 0\;,
\end{equation}
where $c_l$ and $c_t$ the longitudinal wave and transverse wave
velocities, respectively~\cite{white_elastic_1958}. Therefore the
scalar potential is connected to purely longitudinal waves, while the
vector potential is connected to purely transverse (shear)
waves. Moreover for the in-plane waves we consider here it can be
easily shown that $\mathbf{A} (\mathbf{r}) = A_z (\mathbf{r})
\hat{\mathbf{z}}$. These potentials carry the same phase shift
information as the initial fields. However these potentials need not
be continuous along interfaces. We note that the quantity
$\hat{\mathbf{z}} \cdot \frac{1}{2} \nabla \times
\mathbf{u}(\mathbf{r})$, which in the linear elastic regime describes
the angle of shear rotations (transverse part of the field) around the
$z$ axis~\cite{malvern_introduction_1969}, is proportional to $A_z
(\mathbf{r})$, while the quantity $\nabla \cdot
\mathbf{u}(\mathbf{r})$, which express local changes in the volume
(area for 2D) of the structure~\cite{malvern_introduction_1969}, is
proportional to $\phi (\mathbf{r})$.

\begin{figure*}
  \centerline{\includegraphics*[width=\textwidth]{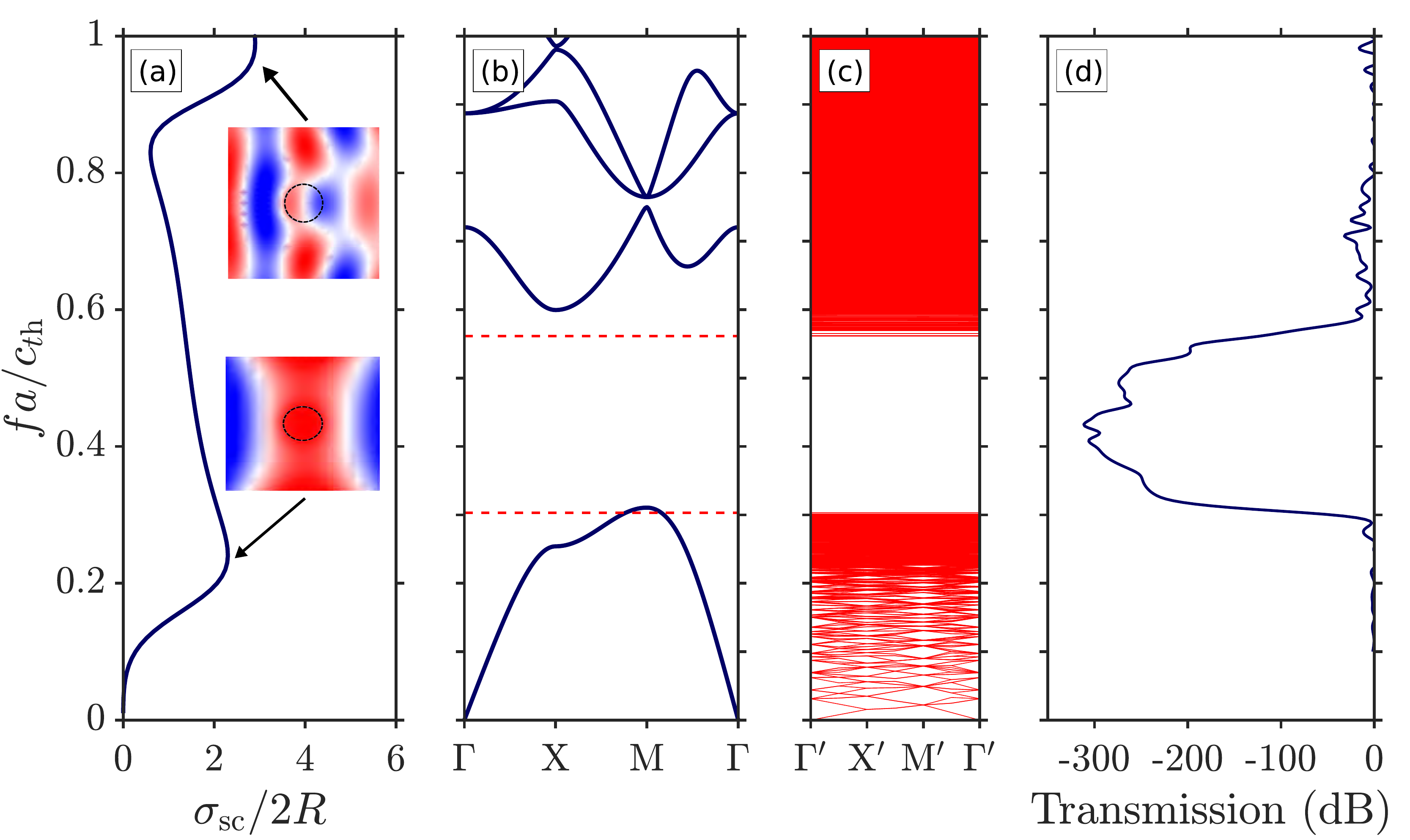}}
  \caption{\label{fig:BandsZ} (Color online) (a) Scattering cross
    section of a single Pb cylinder in epoxy. The field profile at the
    two peaks of the scattering cross section, associated with
    excitation of the first and second eigenmodes of the cylinder are
    shown in the insets (dotted circles mark the position of the
    cylinder). (b): Band structure of the out-of-plane modes of the
    square periodic arrangement with lattice constant $a$.  The band
    gap edges of the HDPS structure are drawn with dashed red lines
    for comparison. (c) Band structure of the out-of-plane modes of
    the HDPS (folded in the supercell). We note that in the long
    wavelength limit the slopes are the same, but look different due
    to the different sizes of the corresponding Brillouin zones. (d)
    Transmission through a finite along the x-direction (length
    $\sqrt{N}a$) slab of the structure.}
\end{figure*}
We consider cylinders made of lead (Pb) with density
$\rho_{\text{Pb}}=11.4$~g/cm$^3$, transverse velocity
$c_{l\text{Pb}}=$2160~m/s, and longitudinal velocity
$c_{t\text{Pb}}=860$~m/s in an epoxy host matrix with
$\rho_{\text{h}}=1.2$~g/cm$^3$, $c_{l\text{h}}=2830$~m/s, and
$c_{t\text{h}}=1160$~m/s, respectively~\cite{chen_study_2007}. The
filling fraction of the structure is $20\%$. These materials and the
filling fraction correspond to the ones used in the literature for a
2D disordered structure, originating from a square
lattice~\cite{chen_study_2007,chen_localisation_2010}, for
comparison. However the results are expected to be similar for other
relatively hard disks in relatively soft embedding solids. Moreover
the filling fraction is not optimized for the disordered case although
the size of the band gaps of the HDPS as compared to the periodic one
suggests that it is close to the optimum value. The phononic band
structure for such a 2D arrangement of cylinders can have two distinct
type of elastic modes: out-of-plane (pure transverse) modes, similar
to 2D TM photonic modes and in-plane (mixed longitudinal and
transverse) modes, corresponding to the 2D TE photonic modes. The band
structure for the out-of-plane modes is shown in the left-hand diagram
of Fig.~\ref{fig:BandsZ}, calculated with a finite element commercial
software (COMSOL Multiphysics\textsuperscript{\textregistered}). The
PBGs for such disordered structures turn out to be equivalent to the
fundamental band gap in periodic systems in the sense that the
spectral location of the TM gap, for example, is determined by the
resonant modes of the individual cylinders (Mie resonances). Indeed,
in the middle diagram of Fig.~\ref{fig:BandsZ}, we see that the
frequency of the lower edge of the band gaps is in a nice agreement
with the first mode of the cylinder, as calculated using
Ref.~\onlinecite{white_elastic_1958}, and in good agreement with
calculations using FEM.

The PBG for the HDPS are equivalent to the fundamental band gap in
periodic systems, since in the former case the band gap occurs between
the $N$, $N+1$ bands, instead of the first and second band of the
periodic structure and we have exactly $N$ scatterers in each unit
(super)cell. This can be interpreted in terms of an effective folding
due to this ``average'' periodicity, which the HDPS exhibits due to
the short-range geometric order. In the disordered (periodic)
structure the PBG extends from $0.30c_{t\text{h}}/a$
($0.31c_{t\text{h}}/a$) to $0.56c_{t\text{h}}/a$
($0.60c_{t\text{h}}/a$), i.e., in the disordered case the PBG
($\Delta\omega/\omega_G = 60\%$) has almost the same size as the
periodic one ($\Delta\omega/\omega_G = 64\%$). However there is a
large number of edge states (localized modes) on both sides of the PBG
in the disordered case, as was also the case for pure random 2D
phononic structures~\cite{chen_localisation_2010}. We note that
frequency scales with length and if we consider, e.g., $a=1$~$\mu$m
then the HDPS PBG extends from 350 to 650~MHz.

The transmission through a finite, along the $x$ axis, slab of the
disordered structure of length $\sqrt{N}a$ is shown in the right-hand
diagram of Fig.~\ref{fig:BandsZ}. We impose periodic boundary
condition along the $y$ axis and perfectly matched layers (PML) along
the $x$ axis. The excitation is done by a uniform load with a
direction along the $z$ axis. We can clearly identify that the transmission
becomes vanishingly small inside the PBG. Moreover, above the PBG the
diffusive propagation of the waves retains relatively small values for
the transmission. This is connected to the observation made by
Sainidou et al.~\cite{sainidou_widening_2005} regarding the
enlargement of the transmission gap in a disordered phononic
structure.

In order to explore the origin of the PBGs in the HDPS we
employ the so-called concentration factor, which in the photonics case
is strongly related to the appearance of band
gaps~\cite{man_photonic_2013}. The concentration factor relates to the
idea that the eigenmodes are minimized states of an energy
functional~\cite{man_photonic_2013}, and can be defined as
\begin{equation}\label{eq:CF}
  C_F = \frac{\int_{\text{rods}}\rho(\mathbf{r}) \left| \mathbf{u}(\mathbf{r}) \right|^2d^2r} 
  {\int_{\text{supercell}}\rho(\mathbf{r}) \left|\mathbf{u}(\mathbf{r}) \right|^2d^2r} \;,
\end{equation}
where $\rho(\mathbf{r})$ is the position-dependent mass density. We note that
$\sqrt{\rho}(\mathbf{r}) \mathbf{u}(\mathbf{r})$ are the eigenmodes of the eigenvalue problem
of the elastic field~\cite{sainidou_greens_2004} and $\rho(\mathbf{r})
|\mathbf{u}(\mathbf{r})|^2$ is proportional to the kinetic energy density. We
expect a PBG to open up when, with increasing energy/frequency, the
modes can no longer spatially distribute in the same manner but have
to drastically change the location or number of nodes, changing
between the two distinct manners of propagation, i.e., elastic wave
hoping coherently from a cylinder to its neighbors (coupled
resonances) and wave propagation mainly through the host
material~\cite{kafesaki_elastic_1995}.

\begin{figure}[h]
  \centerline{\includegraphics*[width=\linewidth]{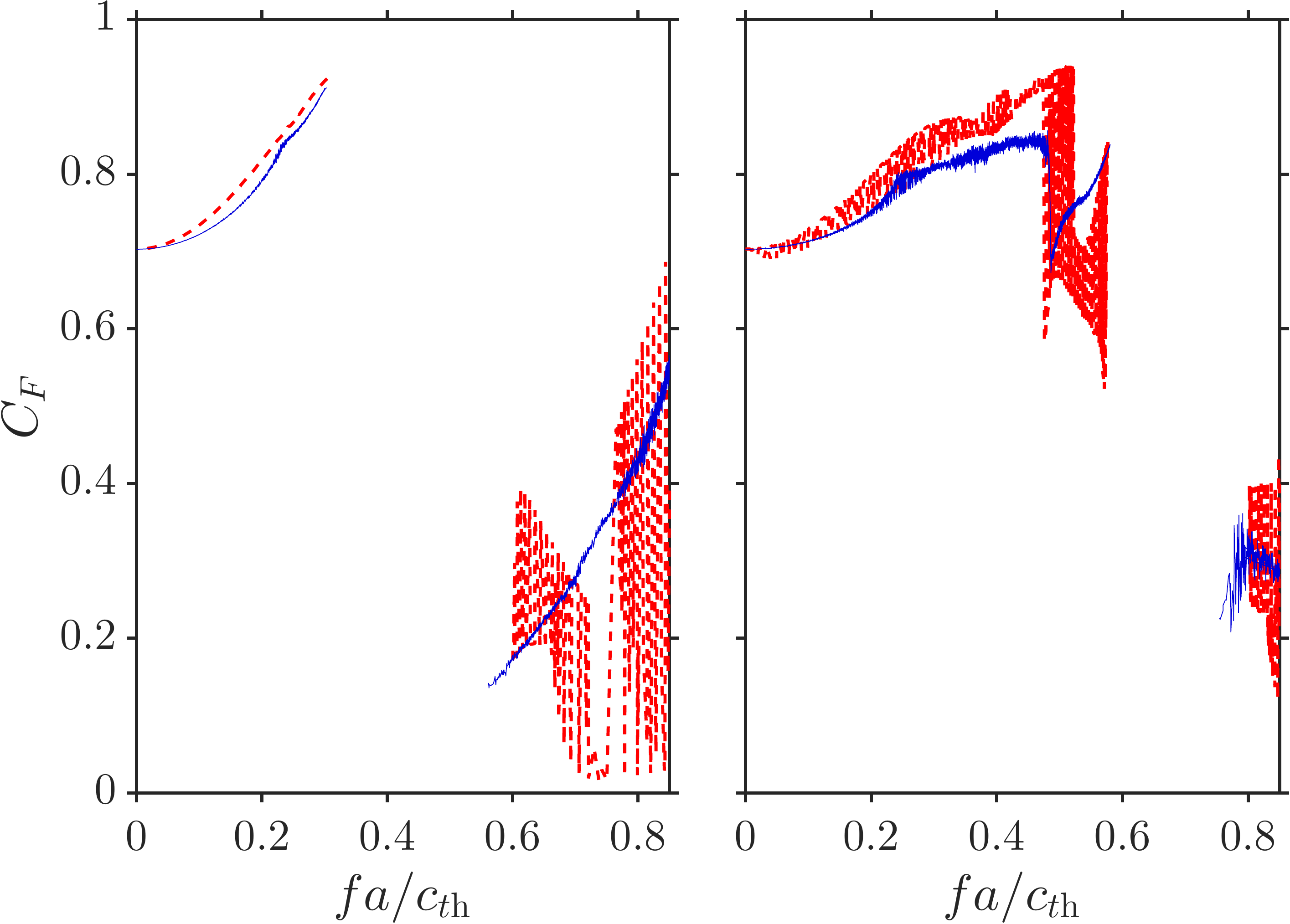}}
  \caption{\label{fig:cFactors} (Color online) The concentration
    factor, $C_F$, as a function of the normalized frequency,
    $fa/c_{t\text{h}}$, for a large number of $\mathbf{k}$ vectors for
    the HDPS (solid black lines) and for the relevant periodic
    structure (dashed red lines). In the left-hand (right-hand)
    diagram the results correspond to the out-of-plane (in-plane)
    modes.}
\end{figure}
Indeed, the left-hand panel of Fig.~\ref{fig:cFactors} shows the
sudden drop of the $C_F$, above the band edge, which implies that
now the kinetic energy is mainly distributed in the host matrix, while
below the PBG it was mainly distributed inside the cylinders. For the
periodic structure there is strong dispersion of the position of the
curves for different $\mathbf{k}$ values which is a direct consequence
of the fact that the periodic structure is an anisotropic structure,
while the HDPS is an isotropic one. Finally, we note a slight anomaly
of $C_F$ at about $0.23c_{t\text{h}}/a$, connected to the change in
the spatial distribution of the modes, as we discuss in the next
paragraph.

\begin{figure}[h]
  \centerline{\includegraphics*[width=\linewidth]{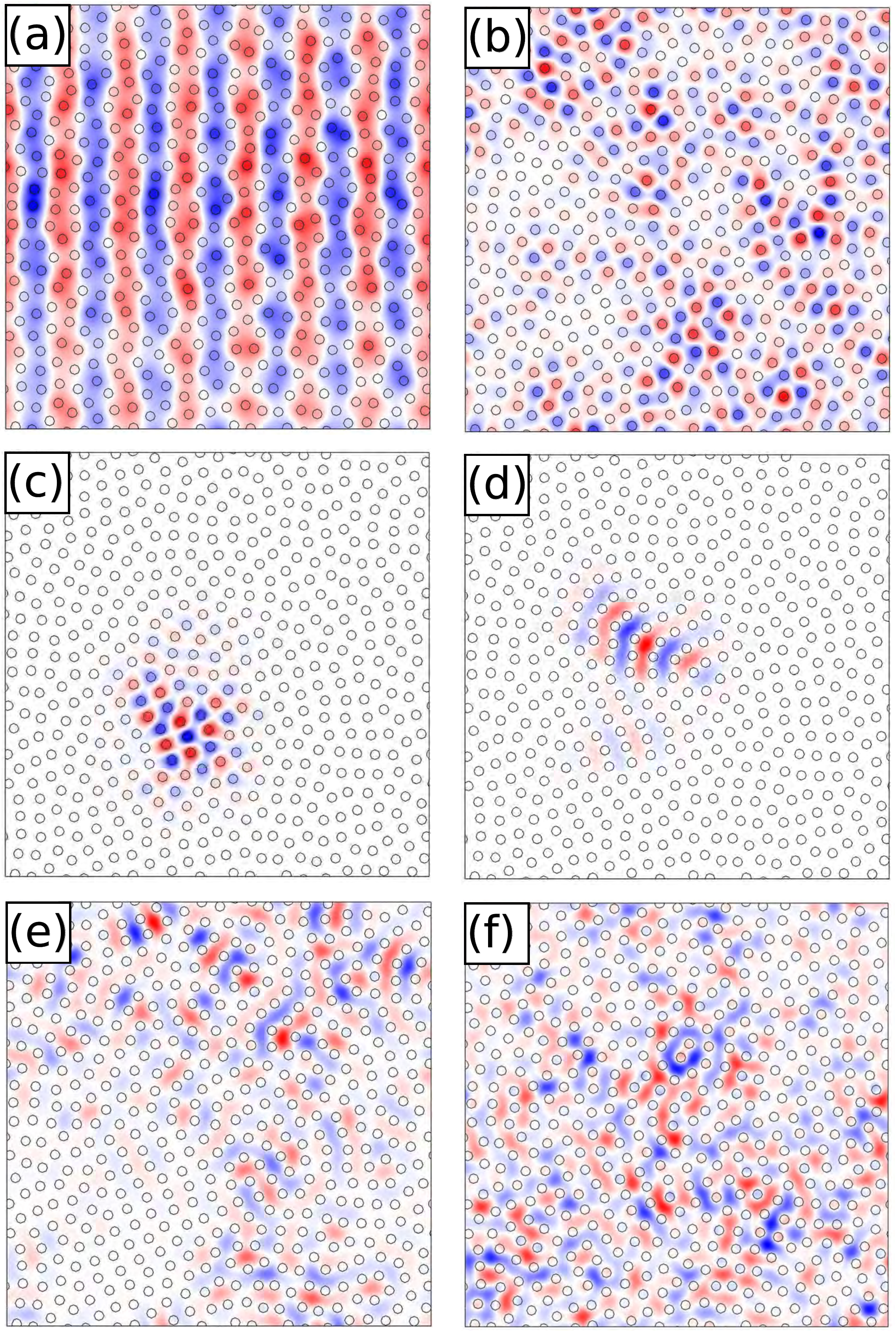}}
  \caption{\label{fig:z-modes} (Color online) $\Re[u_z(\mathbf{r})]$
    for the HDPS at $0.20c_{t\rm{h}}/a$, $0.29c_{t\rm{h}}/a$,
    $0.30c_{t\rm{h}}/a$ (at the band edge), $0.56c_{t\rm{h}}/a$ (at
    the band edge), $0.59c_{t\rm{h}}/a$, $0.65c_{t\rm{h}}/a$, from (a)
    to (f), respectively.}
\end{figure}
The real part of the $z$ component of the elastic displacement field,
$u_z$, is presented in Fig.~\ref{fig:z-modes}. In the long wavelength
limit [Fig.~\ref{fig:z-modes}(a)] the HDPS seems to act as a
homogeneous effective medium, however the inhomogeneity still plays an
important role and we observe strong enhancement of the field on
specific cylinders, or aggregates of cylinders acting as local
scatterers. At higher frequencies, below the band gap, the modes mostly
resemble monopole like modes, with cylindrical symmetry of the
displacement field within the cylinders and the energy concentrated
within the cylinders, suggesting a hoping mechanism for
transmission. We clearly see that as we go pass $0.23c_{t\text{h}}/a$,
i.e from Fig.~\ref{fig:z-modes}(a) to Fig.~\ref{fig:z-modes}(b) the
modes change from collective to more isolated ones, connected to the
slight change in $C_F$. Moreover there is a phase difference of
approximately 180$\degree$ of the field among neighboring
cylinders. We also note that there is a large number of strongly
localized modes in the vicinity of the PBG (edge states). The two
modes in Figs.~\ref{fig:z-modes}(c) and~\ref{fig:z-modes}(d) are such
an example exactly at the band edge. Above the PBG there is increased
concentration of the fields between the scatterers, which supports the
idea that the elastic waves are diffusely transmitted in between the
scatterers. Moreover the modes above the band gap show a dipole-like
behavior, similar to the one of the second resonance of the single
cylinder (see the left-hand diagram of Fig.~\ref{fig:BandsZ}).

\begin{figure*}
  \centerline{\includegraphics*[width=\textwidth]{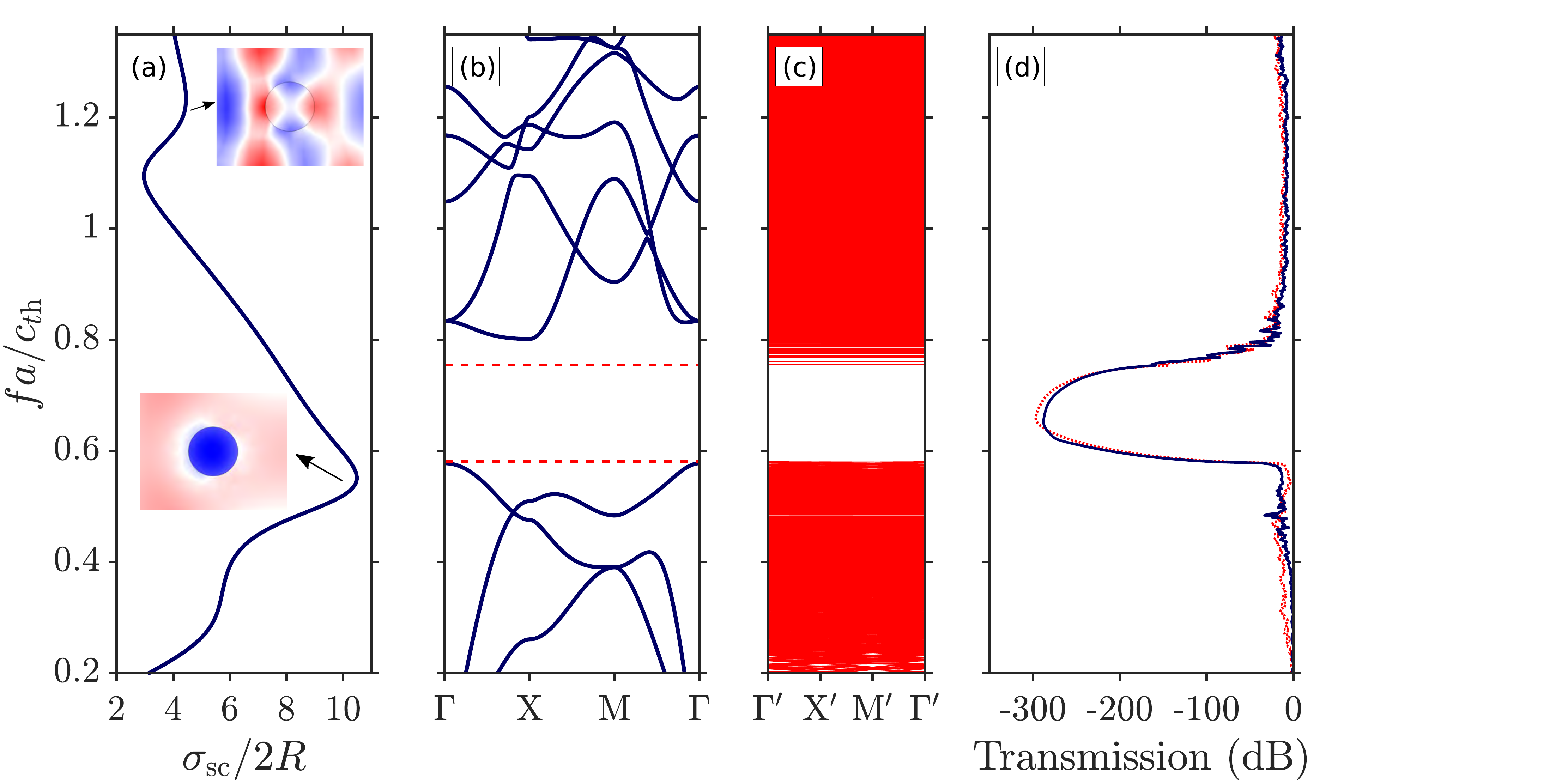}}
  \caption{\label{fig:BandsXY} (Color online) (a) Scattering cross
    section of a single Pb cylinder in epoxy, for shear-polarized
    in-plane incident plane wave. The field profile,
    $\Re[A_z(\mathbf{r})]$, at the two peaks of the scattering cross
    section, associated with the single cylinder modes at
    0.55$c_{th}/a$ and 1.23$c_{th}/a$, respectively, are shown in the
    insets (cylinders can be identified due to the discontinuity of
    these fields). (b) Band structure of the out-of-plane modes of the
    periodic structure with lattice constant $a$. The band gap edges
    of the HDPS structure are drawn with dashed red lines for
    comparison. (c) Band structure of the out-of-plane modes of the
    periodic structure HDPS (folded in the supercell). (d)
    Transmission through a finite along the x-direction (length
    $\sqrt{N}a$) slab of the structure. Transmission is shown for both
    longitudinal (along the x, black solid curve) and the shear (along
    the y axis, red dashed curve) polarizations.}
\end{figure*}
Let us now consider the in-plane modes which, being a mixture of
longitudinal and transverse waves, are expected to have a more complex
behavior. The band structure for these modes is shown in
Fig.~\ref{fig:BandsXY}(c). In Fig.~\ref{fig:BandsXY}(b) we show the
band structure of the corresponding periodic structure with the same
filling fraction. We note that the Brillouin zones in the two cases
are not the same, due to the effective folding of the HDPS. The area
of the HDPS supercell band structure (denoted by prime symbols) is the
$1/N$ of the periodic one. However, in the long wavelength
limit the band structures have the same slopes, as expected for a
homogenized material of the same filling fraction. Moreover, in this
case the opening of the band gap occurs between the bands 1500 and
1501, instead of the third and fourth band of the periodic structure,
indicating again a 500 folding due to the ``average'' periodicity
(short-range order) of the structure.

\begin{figure}[h]
  \centerline{\includegraphics*[width=\linewidth]{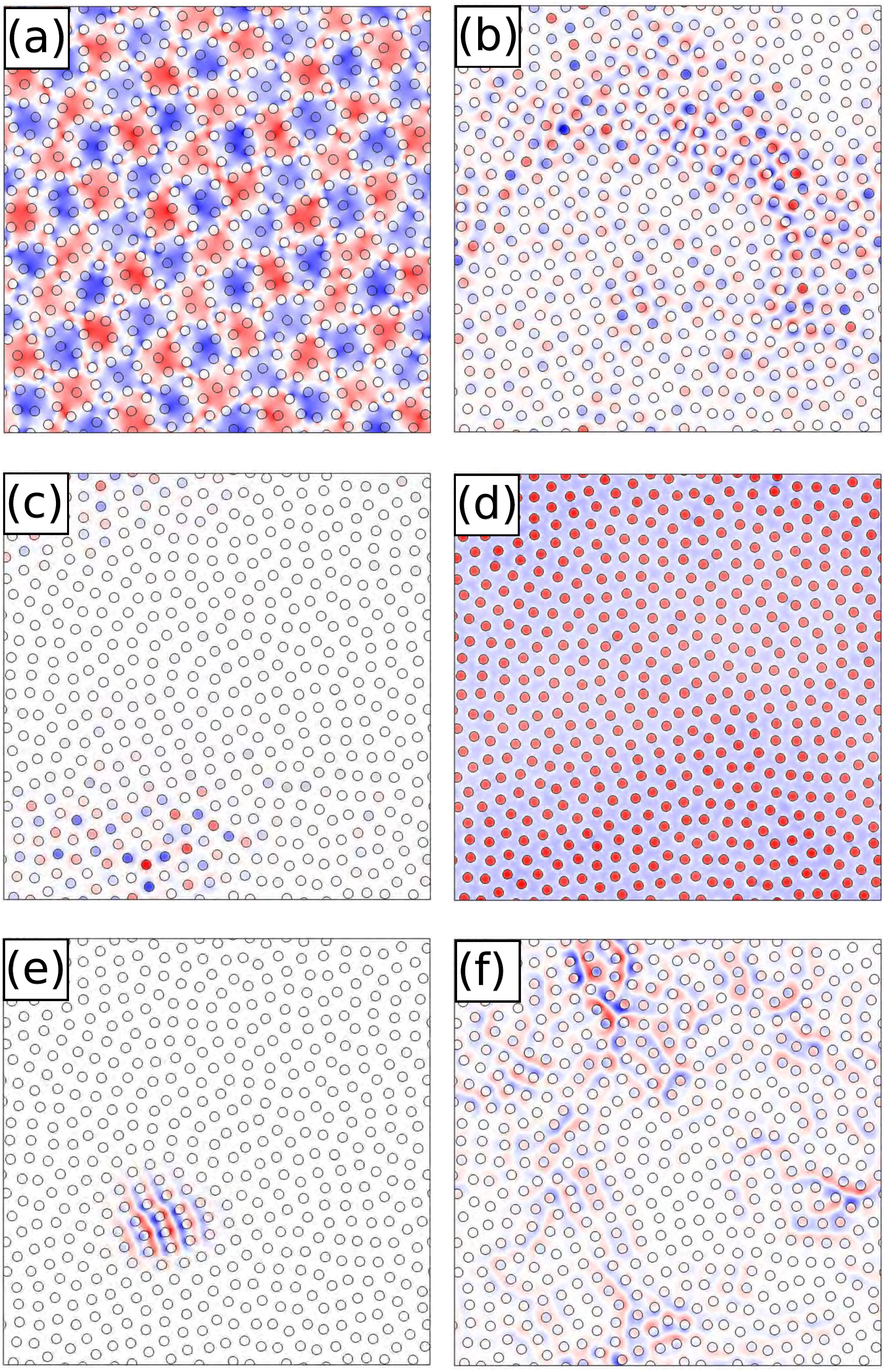}}
  \caption{\label{fig:ModesXY} (Color online) Profile of
    $\Re[A_z(\mathbf{r})]$ for the in-plane modes of the HDPS at
    \textbf{(a)} $0.20c_{t\rm{h}}/a$, \textbf{(b)}
    $0.48c_{t\rm{h}}/a$, \textbf{(c)} $0.49c_{t\rm{h}}/a$,
    \textbf{(d)} $0.58c_{t\rm{h}}/a$ (at the lower edge of the PBG), \textbf{(e)}
    $0.75c_{t\rm{h}}/a$ (at the upper edge of the PBG), \textbf{(f)}
    $0.81c_{t\rm{h}}/a$.}
\end{figure}

However, for the in-plane polarization the HDPS (periodic structure)
band gap extends from $0.58c_{t\text{h}}/a$ ($0.58c_{t\text{h}}/a$) to
$0.75c_{t\text{h}}/a$ ($0.80c_{t\text{h}}/a$), i.e., in the disordered
case the band gap ($\Delta\omega/\omega_G = 26\%$) is smaller than the
periodic one ($\Delta\omega/\omega_G = 32\%$), implying that the
in-plane modes are more sensitive to disorder. Of course, the band gap
is still quite large due to existence of the sort-range order in the
HDPS. We should also note that in this case the structure supports
many edge states (localized modes). Moreover, interestingly enough,
the lower band edge of the HDPS and the periodic structure
coincide. This is not accidental, but has to do with the localized
mode of the single cylinder, shown in the lower inset of
Fig.~\ref{fig:BandsXY}(a), which interacts with the band gap and forms
the lower band edge mode. For this mode see also
Fig.~\ref{fig:ModesXY}(d) and relevant discussion. However, the upper
band edge has shifted to lower frequencies due to the formation of a
large number of localized modes near the top band gap edge. Moreover
the modes above the band gap edge resemble the field profile of the
single mode shown in the upper inset of Fig.~\ref{fig:BandsXY}(a).

The transmission in this case, shown in the right-hand diagram of
Fig.~\ref{fig:BandsXY}, has similar behavior as for the out-of-plane
polarization and is almost independent of the polarization of the
incident wave. We can again see that due to diffusive propagation
transmission is low above the band gap. For the in-plane waves the
drop in the transmission ($-300$~dB) within the band gap is somewhat
smaller compared to the out-of-plane ones ($-320$~dB). These results
indicate that the in-plane modes are more affected by the
disorder. However, overall, we still have sufficiently large band gaps
and corresponding transmission drops.

In the right-hand diagram of Fig.~\ref{fig:cFactors} we show the
concentration factor for the in-plane modes. We again see a similar
behavior with the kinetic energy distribution concentrated inside the
cylinders below the band gap and outside the cylinders above the band
gap. However in this case both the periodic structure and the HDPS
exhibit stronger anisotropy than in the out-of-plane
polarization. Moreover, $C_F$ abruptly drops at about
$0.48c_{t\text{h}}/a$, while remaining almost constant in the region
0.42--$0.49 c_{t\text{h}}/a$, connected to the fact that there is a
decrease in the density of states, as is indicated by the decreased
density in the number of bands we find over this region. Such a
behavior mirrors a similar behavior in
photonics~\cite{tsitrin_unfolding_2015}.

Fig.~\ref{fig:ModesXY}(a) depicts the transverse component of a mode
well below the band gap (long wavelengths). We see collective
oscillations of $Re[A_z(\mathbf{r})]$ corresponding to aggregates of
four to six cylinders. As we approach $0.49 c_{t\text{h}}/a$ the modes
display a mixed character [see Fig.~\ref{fig:ModesXY}(b)]. The
smoothness of the disordering vanishes and local excitation can be
separated into cylindrical symmetry modes within the cylinders and
mirror-symmetric modes with one nodal plane within the cylinders,
resembling dipole-like shaped modes.  The modes are still very
extended. However, just a little bit higher in frequency
[Fig.~\ref{fig:ModesXY}(c)] the modes rapidly localize. Interestingly
whilst doing so they keep their mixed-resonance character. At even
higher frequencies the modes take on a clear monopole-like appearance
and gradually evolve from localized back to extended. Initially the
monopole-like resonances have alternating phase on each inclusion,
however, as maximum localization is surpassed the phase behavior
begins to cluster and eventually fully correlates, just below the
absolute band gap [Fig.~\ref{fig:ModesXY}(d)]. This behavior is quite
peculiar and was not found for the out-of-plane modes. The origin of
this effect has to do with the existence of the highly localized
torsional mode near the band gap edge, which interacts with the band
gap and results in the formation of the band edge mode shown in
Fig.~\ref{fig:ModesXY}(d). In both the periodic and the disordered
case this mode has the same shape and since it is localized in the
cylinders it propagates with a very small group velocity
(~0.001$c_{th}$) through a hoping mechanism. As a result the
geometrical disorder is not affecting strongly this mode. At the upper
band edge the modes exhibit the usual tight localization
[Fig.~\ref{fig:ModesXY}(e)], while at even higher frequencies the
modes become more extended but still highly diffusive
[Fig.~\ref{fig:ModesXY}(f)].

There are some similarities to the photonics case in the sense that we
observe the displacement field localized mainly inside, below the band
gap, and outside the cylinders, above the band gap. Moreover we see
that the ballistic transport, at low wavelengths, is acompanied with
excitation of local modes in the cylinders, while we see diffusive
transport well above the band gap. However, the modes also show
important differences. Firstly, we see in general a higher
localization of the modes. Secondly, a delocalized band
edge mode, with a very low group velocity ~0.001$c_\mathrm{th}$, was
identified in the lower edge of the band gap, due to the existence of
a torsional mode at the lower edge of the band gap. Such an extended
mode was not observed previously in hyperuniform photonic
structure. Thirdly, the modes above the band gap remain localized for
a large frequency region, up to about 0.81$c_\mathrm{th}$, before
turning to diffusive modes. We note that these results could in
principle be found also in the photonics case for appropriate
combination of material parameters.

\begin{figure}[h]
  \centerline{\includegraphics*[width=\linewidth]{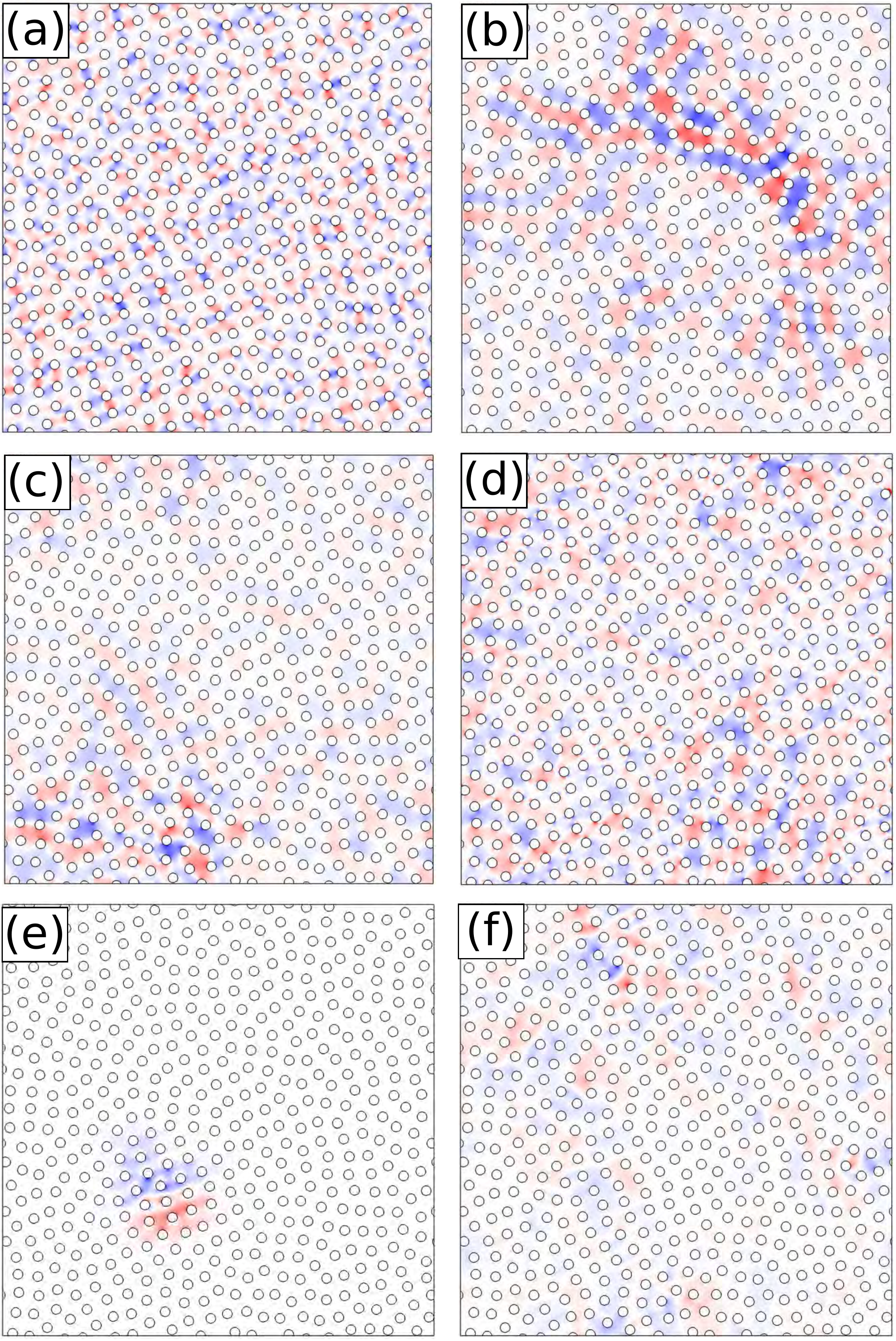}}
  \caption{\label{fig:ModesXY_Div} (Color online) Profile of
    $\Re[\phi(\mathbf{r})]$ for the in-plane modes of
    the HDPS at \textbf{(a)} $0.20c_{t\rm{h}}/a$, \textbf{(b)}
    $0.48c_{t\rm{h}}/a$, \textbf{(c)} $0.49c_{t\rm{h}}/a$,
    \textbf{(d)} $0.58c_{t\rm{h}}/a$ (at the band edge), \textbf{(e)}
    $0.75c_{t\rm{h}}/a$ (at the band edge), \textbf{(f)}
    $0.81c_{t\rm{h}}/a$.}
\end{figure}

The longitudinal component of the fields is shown in
Fig.~\ref{fig:ModesXY_Div}. In contrast to the transverse components
the longitudinal ones exhibit always a stronger delocalization and no
obvious correlations. However at high frequencies [see, e.g.,
Fig.~\ref{fig:ModesXY_Div}(e,f)] the mode profiles of the longitudinal
and the transverse components become quite similar, showing the
striking behavior of spreading out continuously along curvy planes
among the scatterers.

\section{\label{sec:CavWG}Cavities and waveguides}

A cavity in a HDPS can be created by simply reducing the radius of a
single cylinder or totally removing it. In the photonics case it was
observed that the localized modes have some resemblance to the modes
at the band edge. We expect some differences to the cavity and
waveguide modes of photonic band gap structures, due to the different
mode behavior around the band gap for the in-plane modes of the
phononic structure.

\begin{figure}[h]
  \centerline{\includegraphics*[width=0.5\linewidth]{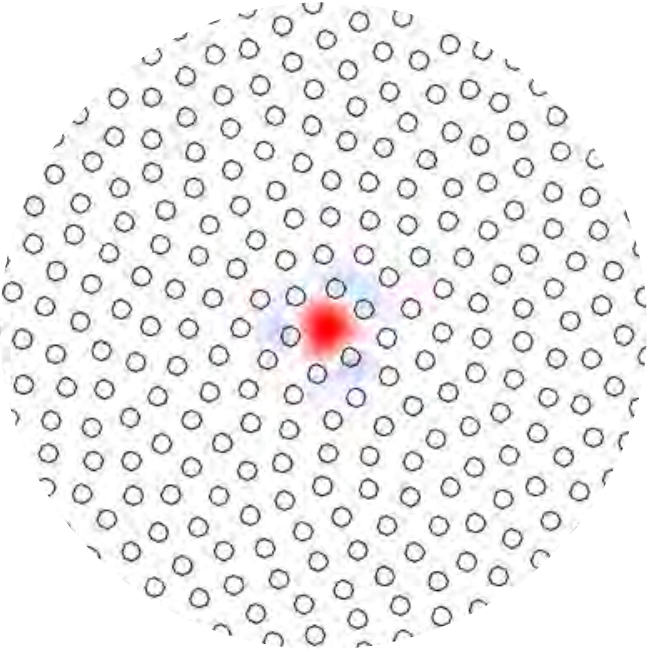}}
  \caption{\label{fig:Cav_Z} (Color online) The $u_z$ cavity mode
    profile for the in-plane polarization. The frequency of the
    eigenmode at $0.43c_{t\rm{h}}/a$ has a $Q$ factor of
    $5\times10^{15}$.}
\end{figure}
Fig.~\ref{fig:Cav_Z} shows the out-of-plane resonant cavity mode
introduced by removing a cylinder from the HDPS. In this case we
consider a finite number of 499 cylinders and substitute the periodic
boundary conditions with PML. The cavity mode, which has a cylindrical
symmetry, occurs at $0.43c_{t\rm{h}}/a$ and has a very high $Q$-factor
of about $5\times10^{15}$. A careful look in the mode profile also
reveals the flexibility of the mode to exactly adjust its shape around
the surrounding cylinders. We note that this value is much larger than
what is usually found in the literature on phononic cavities. The
reason for that is that we do not take into account phonon-phonon
scattering which limits the intrinsic Q-factor. This intrinsic
Q-factor is usually of the order of $10^5$ or less depending on
specific frequencies and materials considered\cite{_realizing_2010}. As a result usually a
limited number of layers is enough to achieve this intrinsic Q-factor
while in our case there are about 11 layers surrounding the
cavity. Therefore in all cases considered we are well above that point
and only intrinsic Q-factor limits should be considered on specific
applications.

\begin{figure}[h]
  \centerline{\includegraphics*[width=\linewidth]{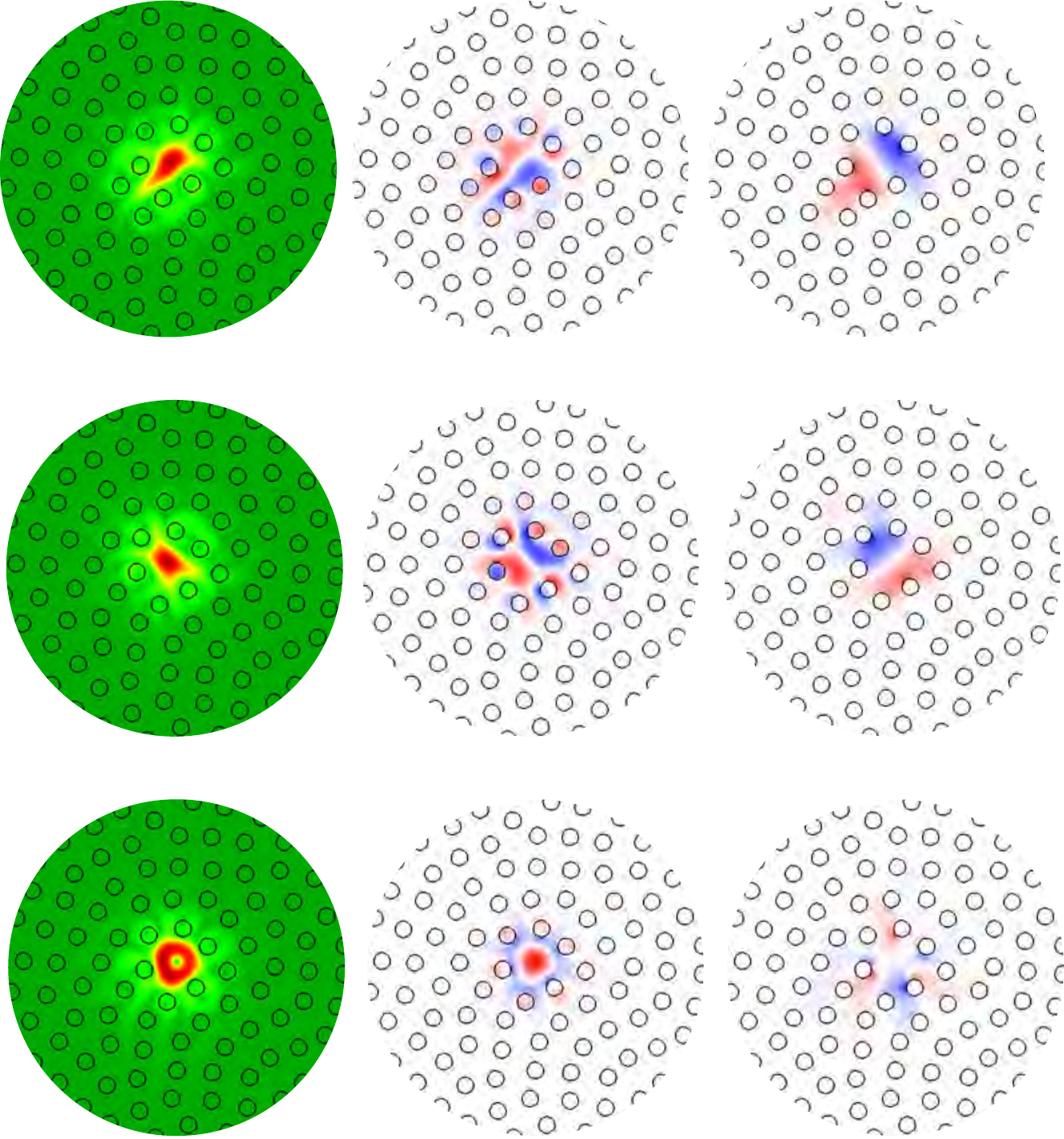}}
  \caption{\label{fig:Cav_XY} (Color online)
    $|\mathbf{u}(\mathbf{r})|$ (left-hand column),
    $\Re[A_z(\mathbf{r})]$ (middle column), and $\Re[\phi
    (\mathbf{r})]$ (right-hand column) cavity mode profiles for the
    in-plane polarization. The frequencies of the eigenmodes at
    $0.66c_{t\rm{h}}/a$ (top row), $0.67c_{t\rm{h}}/a$ (middle row)
    and $0.71c_{t\rm{h}}/a$ (bottom row) have $Q$ factors
    $9\times10^{13}$, $2\times10^{14}$, and $3\times10^{13}$,
    respectively.}
\end{figure}
Fig.~\ref{fig:Cav_XY} shows the in-plane resonant modes of the
cavity. In this case three cavity modes are formed. The first two
modes, at $0.66c_{t\rm{h}}/a$ ($Q=9\times10^{13}$) and
$0.67c_{t\rm{h}}/a$ ($Q=2\times10^{14}$), appear elongated with a
plane of nearly mirror symmetry, almost normal between them. The
lower-frequency more spread out in the lateral direction by
approximately $2a$. The higher-frequency mode is more tightly confined
along the symmetry direction and its shape appears more strongly
influenced by the adjacent scatterers. The highest-frequency mode at
$0.71c_{t\rm{h}}/a$ with $Q=3\times10^{13}$ has a nearly cylindrical
symmetry.

In phononic crystals, removing a row of cylinders generates a channel
through which elastic waves with frequencies within the PBG can
propagate, i.e., waveguide modes. Elastic waves cannot propagate
elsewhere in the structure outside the channel because there are no
elastic modes to couple to. However, the waveguides must be composed
of segments whose orientation is confined to the high-symmetry
directions of the crystal. As a result, the waveguide bends of
$60\degree$ or $90\degree$ can be easily
achieved~\cite{khelif_guiding_2004}, but bends at an arbitrary angle
lead to significant scattering losses due to excessively strong
scattering at the bend junction. However in the case of the HDPS the
distribution of cylinders around the bend junction are statistically
isotropic. If the defect mode created by the removal of cylinders
falls within the PBG, the bend can then be oriented at an arbitrary
angle.

\begin{figure}[h]
  \centerline{\includegraphics*[width=\linewidth]{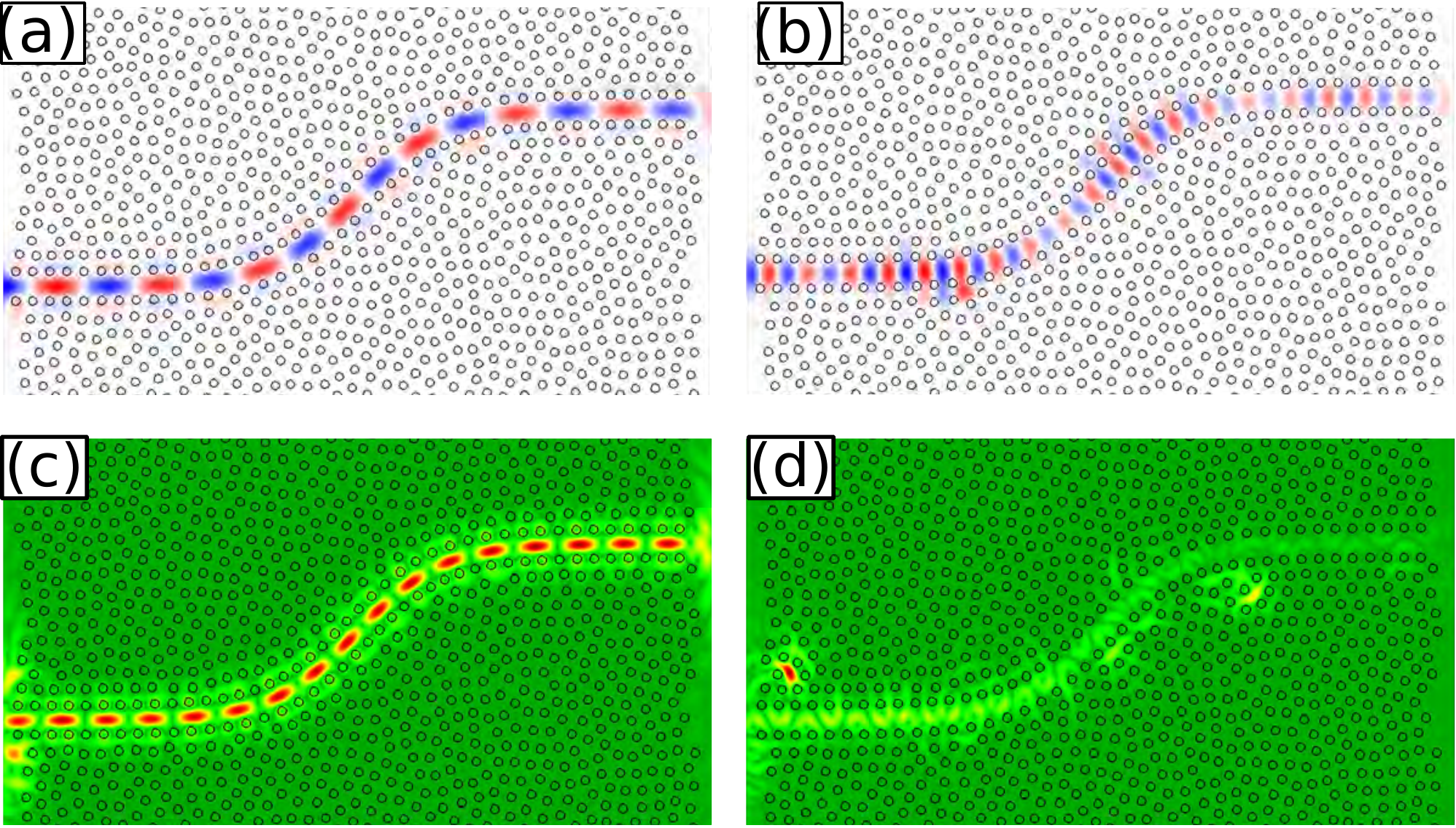}}
  \caption{\label{fig:WGFields} (Color online) $\Re[u_z(\mathbf{r})]$
    for the out-of-plane polarized field for CW excitation at
    $0.41c_{t\text{h}}/a$~(a) and
    $0.54c_{t\text{h}}/a$~(b). $|\mathbf{u}(\mathbf{r})|$ for the
    in-plane polarized field for CW excitation at
    $0.61c_{t\text{h}}/a$~(c) and $0.77c_{t\text{h}}/a$~(d).}
\end{figure}

We note that we could couple such high-$Q$ cavity modes to create a
line of defects and thereof coupled-cavity waveguide
modes~\cite{khelif_trapping_2003}. However we employ here a more
efficient and flexible bottom-up design strategy previously introduced
for photonic structures~\cite{amoah_hyperuniform_2015}. We define the
path of the waveguide first and then built the structure around
it. Specifically, we distribute periodically cylinders on curves
parallel to the waveguide line and then distribute the rest of the
scatterers in a nearly isotropic and homogeneous manner around the
waveguide. We use one layer of ordered cylinders to shape a waveguide
with a sharp bend (as shown in Fig.~\ref{fig:WGFields}) and leave the
rest of the structure disordered. It is worth-noting that in the
photonics case it is not enough to consider only one layer of ordered
cylinders, but at least three. We think the main reason for this is
the fact that phonons are strongly localized around defects, as we can
see from Figs.~\ref{fig:Cav_Z} and ~\ref{fig:Cav_XY}.

\begin{figure}[h]
  \centerline{\includegraphics*[width=\linewidth]{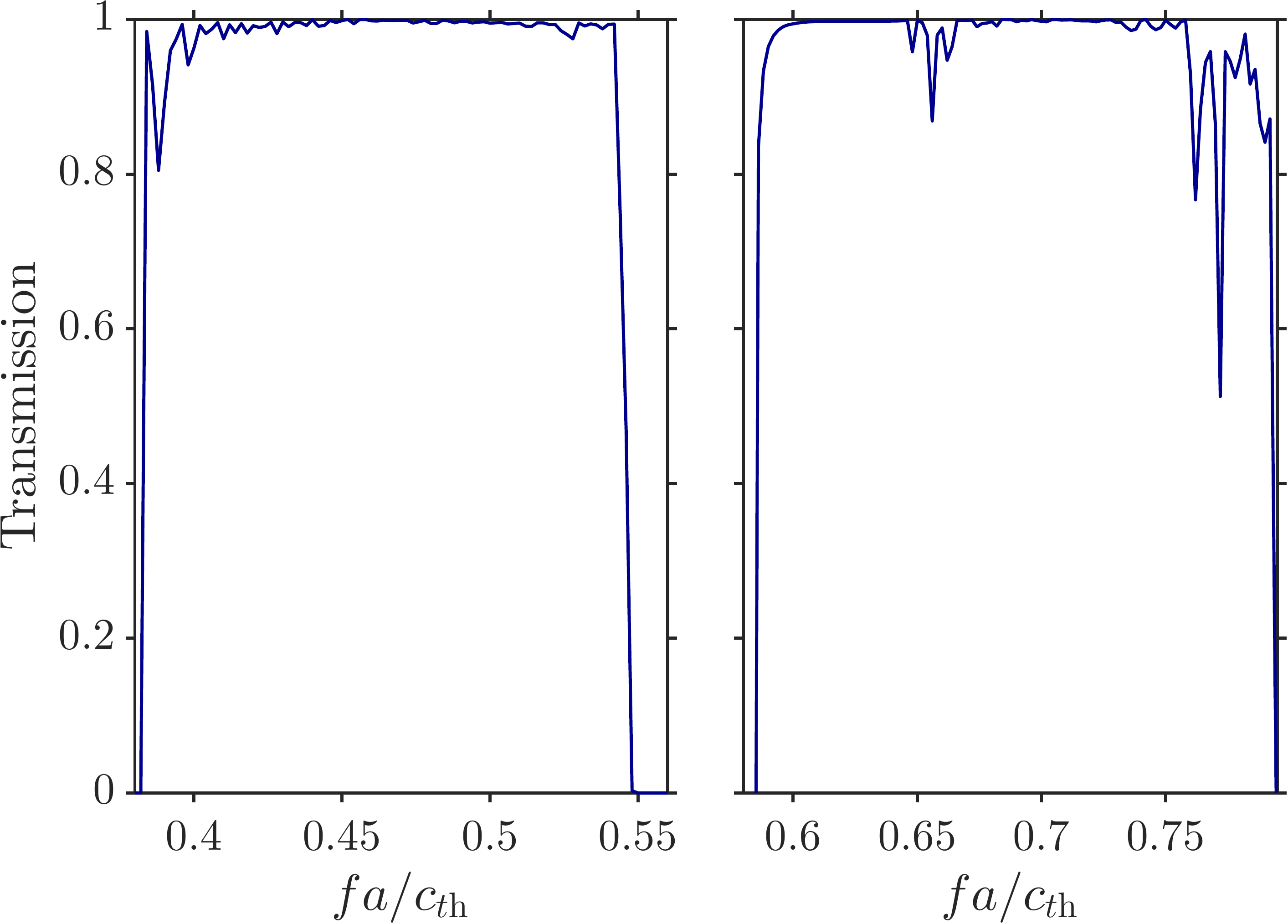}}
  \caption{\label{fig:WG} Transmission spectrum through a curved
    waveguide formed by surrounding HDPS-like structures. Left-hand
    panel: Transmission for out-of-plane polarized waves incident from
    the left (see Fig.~\ref{fig:WGFields}(a)). Right-hand panel:
    Transmission for in-plane polarized waves incident from the left
    (see Fig.~\ref{fig:WGFields}(c)).}
\end{figure}
The left-hand diagram of Fig.~\ref{fig:WG} shows the transmission
spectrum through the waveguide for out-of-plane polarization. In order
to trigger efficiently the transmission through the waveguide we
consider a Gaussian load profile centered at the waveguide region,
with a width of $2a$. We see an almost $100\%$ transmission for this
structure within the corresponding HDPS band gap. For the in-plane
polarization (see right-hand diagram of Fig.~\ref{fig:WG}) we again
observe efficient waveguiding of about $100\%$ transmission and very
strong confinement of the mode within the frequency region of the
corresponding HDPS band gap. However there are some significant drops
in the transmission around $0.65c_{t\text{h}}/a$ and
$0.75c_{t\text{h}}/a$. This is probably due to multimode behavior in
that region, but we do not intent to further investigate such details
in the current manuscript.

In Figs.~\ref{fig:WGFields}(a) and (b) we show the $u_z$ field profile
for continuous wave (CW) excitation normal to the plane at $0.41c_{t\text{h}}/a$, well
inside the waveguide frequency region, and at $0.54c_{t\text{h}}/a$,
at the edge of the waveguide frequency region. It is clear that the
field is highly localized, although in the latter case there are
scattering losses and the field attenuates along the waveguide. As
expected in the second case the elastic field has a much smaller
wavelength, as we can identify by the phase shift. In
Figs.~\ref{fig:WGFields}(c) and (d), we show the magnitude of the
in-plane polarized elastic field at $0.61c_{t\text{h}}/a$, where we
have a high transmission of almost $100\%$ and at
$0.77c_{t\text{h}}/a$, at the strong drop of the transmission (see
Fig.~\ref{fig:WG}). In the former case the field is highly localized
along the waveguide. However in the latter case the field has strong
scattering (diffusive) losses inside the HPDS-like material.

\begin{figure}[h]
  \centerline{\includegraphics*[width=\linewidth]{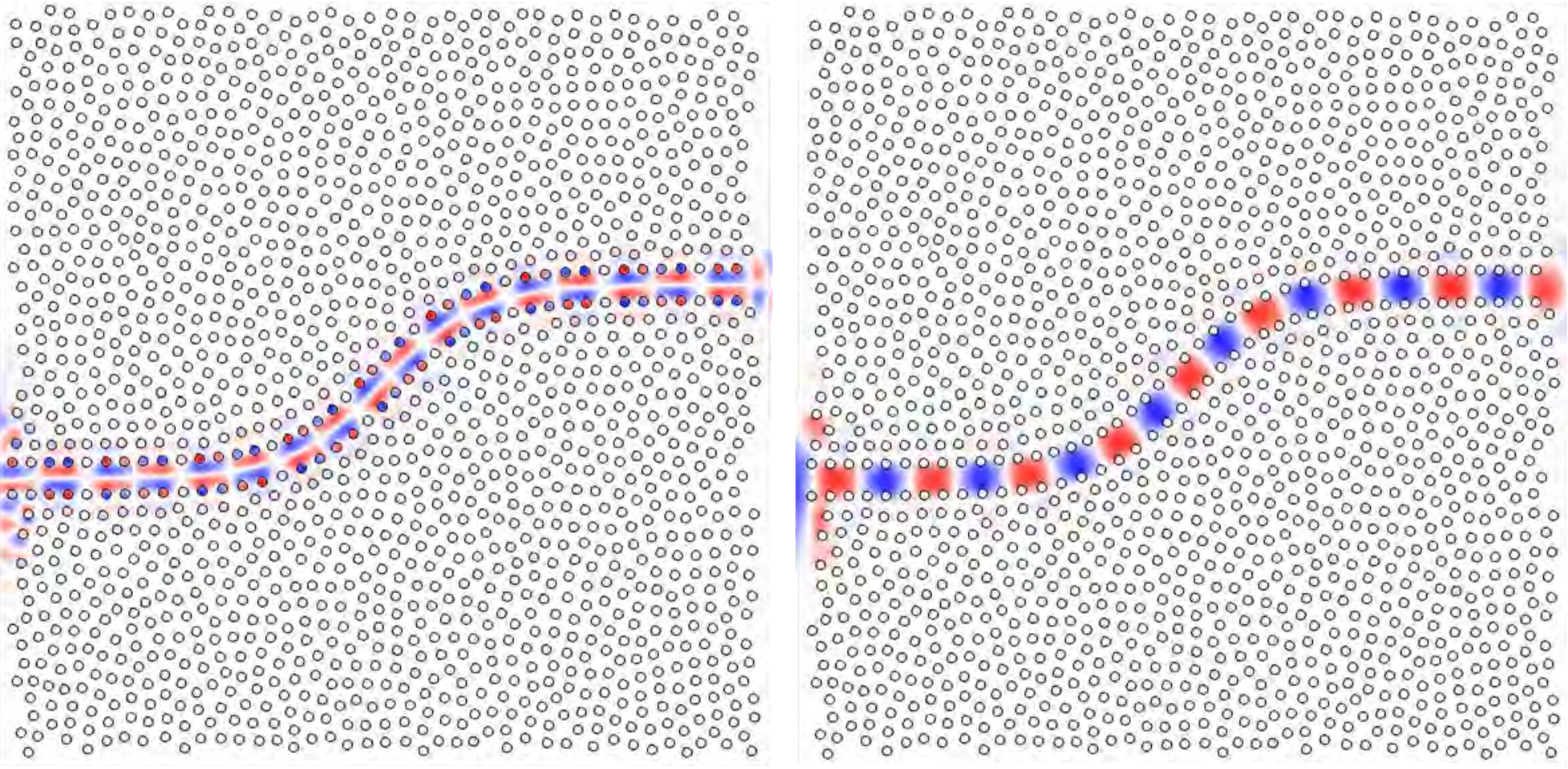}}
  \caption{\label{fig:WG_XY} (Color online) Decomposition of the
    in-plane elastic field of Fig.~\ref{fig:WGFields}(c) in the
    $\Re[A_z(\mathbf{r})]$ (left-hand diagram) and
    $\Re[\phi(\mathbf{r})]$ (right-hand diagram) parts.}
\end{figure}
In order to further understand the behavior of the in-plane field, we
present the decomposition of the in-plane polarized elastic field at
$0.61c_{t\text{h}}/a$ [Fig.~\ref{fig:WGFields}(c)] in
$\Re[A_z(\mathbf{r})]$ (transverse component) and $\phi (\mathbf{r})$
(longitudinal component) in the left- and right-hand diagram of
Fig~\ref{fig:WG_XY}, respectively. We can identify the wavelength from
the sign change of the components along the waveguide. Moreover the
longitudinal component is somewhat better localized within the
waveguide region, while the transverse one expands in the region of
the ordered arrays of cylinders. And this is the small difference in
the strength of the localization among the out-of-plane
[Fig.~\ref{fig:WGFields}(a)] and the in-plane
[Fig.~\ref{fig:WGFields}(c)] polarized elastic fields.

\section{\label{sec:3D}Finite thickness HDPS}
We finally consider a thin finite slab of the HDPS of thickness $a$
along the $z$ axis. In this case we expect Lamb waves, drastically
modulated by the HDPS inhomogeneities, to propagate through the
structure. We note that for such structures the in-plane and
out-of-plane modes are no longer decoupled, but interact with each
other. We consider that the HDPS has a finite length of $\sqrt{N}a$
along the $x$ axis and use PML to account for propagating modes that
travel along the $x$ axis in the (infinite) surrounding epoxy
material. On, the other hand we still consider periodic boundary
conditions for the $y$ axis. Finally we consider stress free top and
bottom surfaces ($z$ axis).

\begin{figure}[h]
  \centerline{\includegraphics*[width=\linewidth]{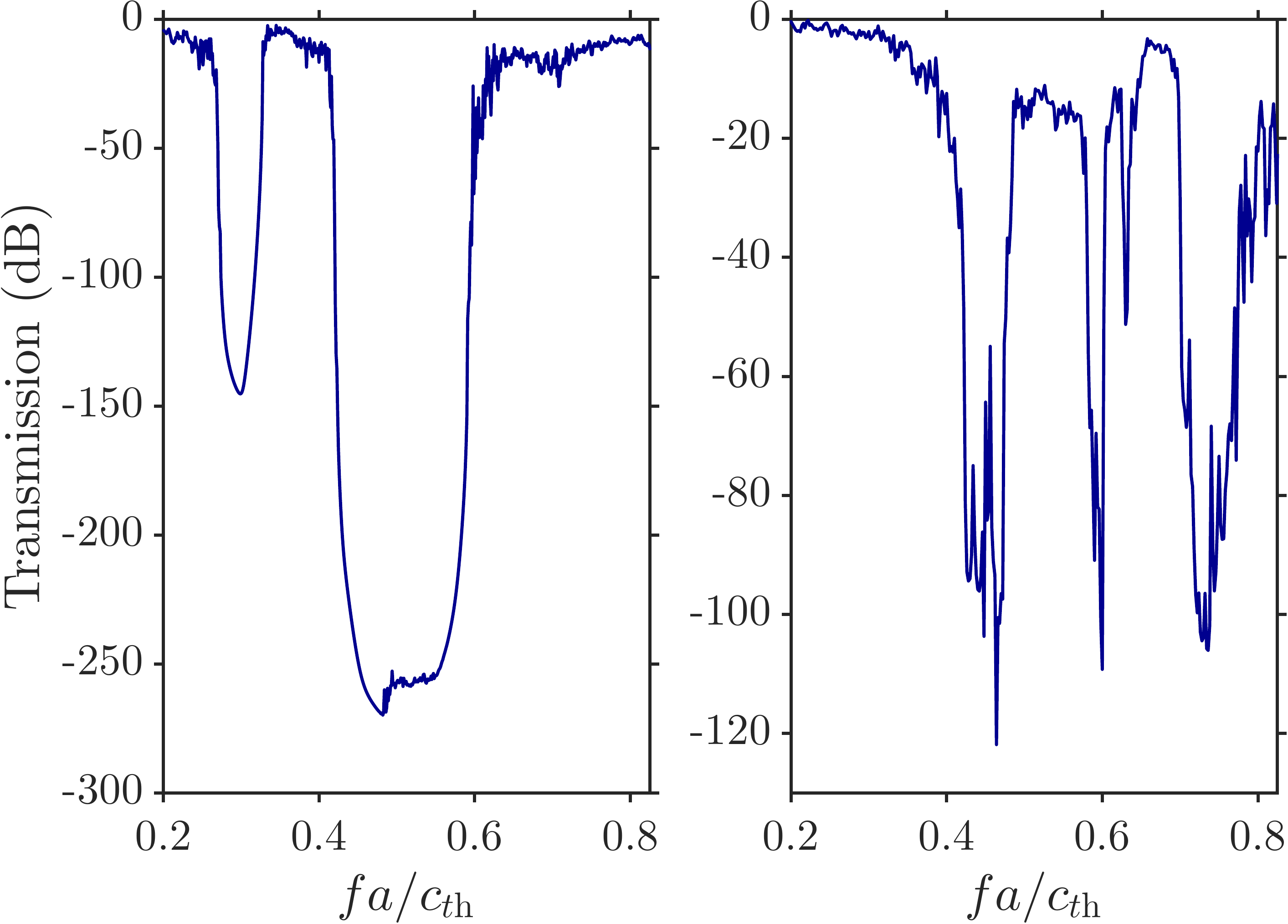}}
  \caption{\label{fig:LambTrans} Transmission through a thin film of
    the HDPS of thickness $a$. The transmission for a boundary uniform
    load applied on the left boundary with a direction along the
    $z$-axis ($x$-axis) is shown in the left-hand (right-hand)
    diagram.}
\end{figure}
The left-hand diagram of Fig.~\ref{fig:LambTrans} shows the
transmission through this structure for an excitation along the $z$
axis. We can clearly identify the large transmission drops within the
corresponding PBGs (see left-hand diagrams of Figs.~\ref{fig:BandsZ}
and~\ref{fig:BandsXY}). However in this 3D case there is a large
number of localized modes within this region, as we can identify from
the transmission peaks. Therefore no absolute band gap exists in this
case. We also note that, independently of the load polarization, we
can see significant transmission drops corresponding to the frequency
regions of the 2D PBGs of both the in-plane and out-of-plane modes,
since all the modes are now coupled. However the transmission gaps are
somewhat shifted, as a result of the finite thickness of the HDPS and
the need for the modes to redistribute their energy profiles in a
finite extent along the $z$ axis (modulated Lamb waves). When we
excite this 3D structure with a uniform load along the $x$ axis most
of the localized modes are not excited and we get a clearer picture
within the transmission band gaps.

\begin{figure}[h]
  \centerline{\includegraphics*[width=\linewidth]{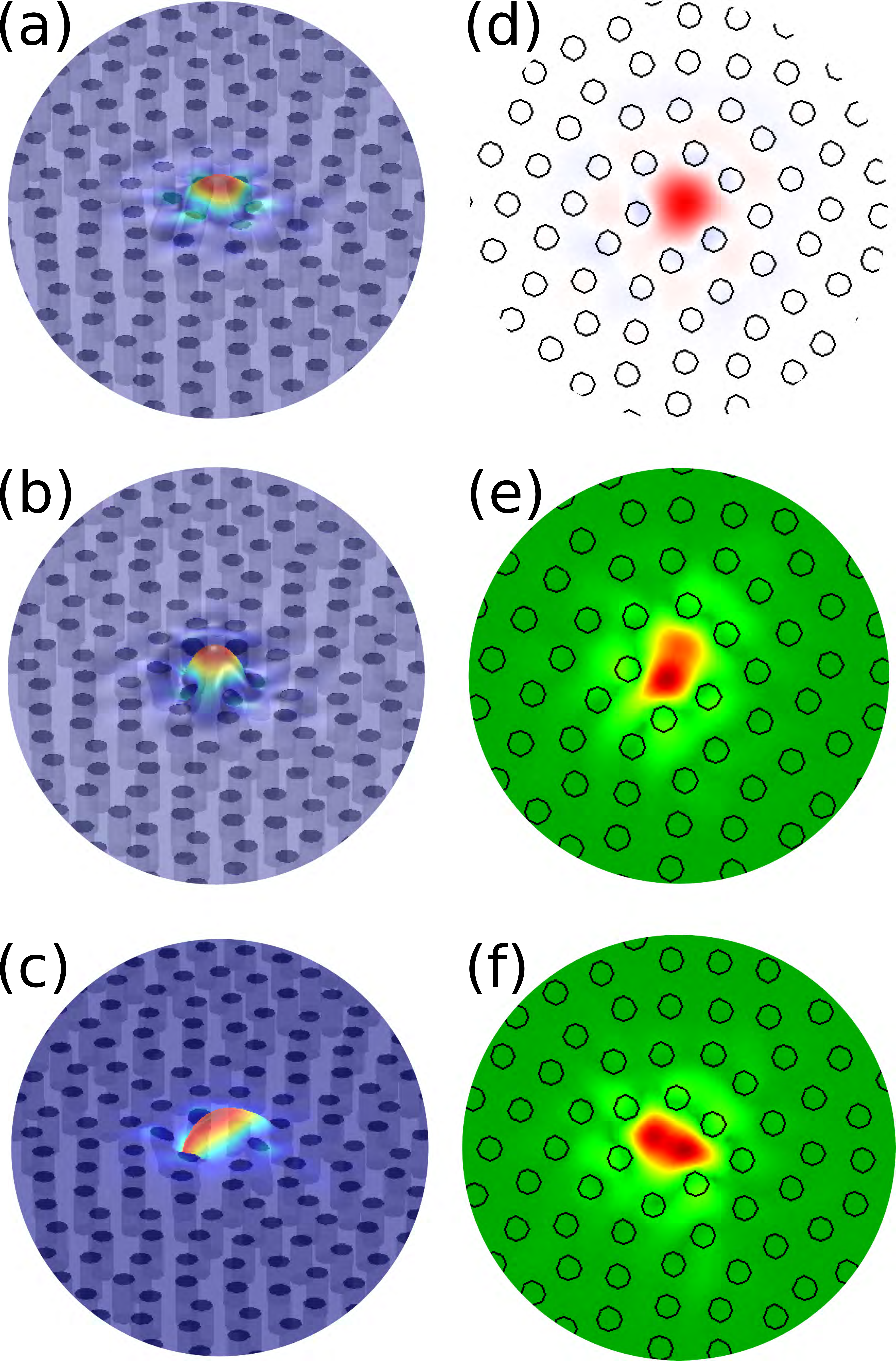}}
  \caption{\label{fig:Cav_3D} (Color online) (a-c) The 3D cavity mode
    profiles for a finite height structure extending from $z=-a/2$ to
    $z=a/2$ at 0.46, 0.66, and 0.67$c_{t\text{h}}/a$ from top to
    bottom, respectively. The deformation of the structure is a
    magnification of the actual deformation. (d) The $u_z$
    displacement profile at $z=0$ for the mode in (a).  (e, f) The
    $|\mathbf{u}|$ displacement profile at $z=0$ for the modes (b) and
    (c), respectively.}
\end{figure}
Moreover, we expect the finite thickness of the HDPS to affect
strongly the cavity modes. As we already mentioned in this case there
is no absolute band gap and no characterization of the modes according
to their polarization. Therefore, even in the large transmission dips
we were able to identify a large number of localized modes. However
among them the ones corresponding to the 2D cavity modes are easily
identified from their very high $Q$ factors. Specifically, in
Fig.~\ref{fig:Cav_3D} we show the modes with high $Q$ factors. We can
identify three such modes. The first, at $0.46c_{t\rm{h}}/a$
[Fig.~\ref{fig:Cav_3D}(a)] with $Q=6\times10^{12}$, originates from
the out-of-plane mode at $0.43c_{t\rm{h}}/a$ ($Q=5\times10^{15}$), as
we can verify by direct comparison of Fig.~\ref{fig:Cav_3D}(d) with
Fig.~\ref{fig:Cav_Z}. The modes at $0.56c_{t\rm{h}}/a$
($Q=1\times10^{14}$) and $0.57c_{t\rm{h}}/a$ ($Q=2\times10^{13}$)
resemble the elongated displacement profiles of the modes at
$0.66c_{t\rm{h}}/a$ ($Q=9\times10^{13}$) and $0.67c_{t\rm{h}}/a$
($Q=2\times10^{14}$) [see Fig.~\ref{fig:Cav_XY}(a,b)], although there
is significant change in the displacement profiles.

\section{\label{sec:conc}Conclusions}

We introduced a new class of structurally disordered phononic
crystals, hyperuniform disordered phononic structures (HDPS). These
structures are made from initially arbitrary point patterns by
applying strong correlations among the points and finally decorating
them with a specific pattern, so that the structure factor becomes
isotropic and vanishes for all $k$-vectors within a specific
radius. Such structures can be clearly identified (see
Fig.~\ref{fig:SuperCell}) as amorphous phononic structures. Although
we only consider here cylindrical inclusions we expect similar results
independently of the shape of the scatterers, and even for more complex
decorations, such as continuous networks originating, similarly to the
photonics case~\cite{florescu_designer_2009}. In such structures both
ballistic propagation (at lower wavelengths) and diffusive transport
(at higher frequencies) together with large phononic band gaps can
coexist.

By using finite element method calculations and super cell techniques
the band structure of HPDS has been calculated. Large PBGs, similar to
the periodic counterparts, have been identified. Such large band gaps
cannot be found in completely random phononic structures. Moreover a large
number of localized modes was found within the frequency region of
interest. As a result, huge transmission drops have been observed in
the frequency regions of the corresponding band gaps, for all
polarizations. These results indicate that these structures can be used
as frequency filters, with the added advantage of being highly
isotropic. This latter characteristic makes them also strong
candidates for in- and out-coupling of free-space elastic waves with
waveguides.

High-$Q$ cavity modes can be, therefore, easily implemented. We showed
here an example of a high-$Q$ cavity by removing a single
cylinder. However, we expect this to be the case even by just altering
the size or misplace one or more cylinders, depending on the mode
localization we want to achieve. We were able to find extremely large
Q-factors, ignoring intrinsic phonon-phonon scattering, and strong
localization of the field profile. Combined with the fact that in
these structures diffusive propagation and a large number of localized
modes are observed, such structures can be efficiently used for
thermal conductivity
reduction~\cite{maldovan_sound_2013,davis_nanophononic_2014,maire_thermal_2015}.
Moreover the fact that these structures are also very efficient in
photonics applications implies that they can be also used to create
optomechanical and/or phoxonic
devices~\cite{eichenfield_optomechanical_2009,aspelmeyer_cavity_2014,lucklum_phoxonic_2013,djafari-rouhani_phoxonic_2016}.
Waveguiding in such structures can also be very efficient. It is shown
that 100\% transmission through arbitrary bends is possible, due to
isotropy. This is a very important result, since this behavior cannot
been found in conventional periodic structures, where you need to
follow high symmetry lines. Therefore such structures are very strong
candidates for phononic integrated
circuits~\cite{iii_microfabricated_2009}.

We have also shown that 3D thin layer HDPS behave similar to the 2D
ones, in terms of the large effective band gaps and the existence of
high-$Q$ cavity modes. This is an important aspect for practical
applications, since very thick (2D) structures are, in many cases, not
efficient. Although we mainly discuss micro and nano-scale
applications we expect that the understanding and control of such
coupling between random and short-range ordered structures is important
in the macroscale as well, such as in the understanding of seismic
wave propagation, fracture in complex structures, as well as
bio-materials, which all include some kind of correlated disorder.

\textbf{Data availability.} The data underlying the findings of this
study are available without restriction. Details of the data and how
to request access are available from the University of Surrey
publications repository: \url{https://doi.org/10.15126/surreydata.00813700}.

\begin{acknowledgments}
  This work was partially supported by the University of Surrey's FRSF
  and IAA awards to MF, EPSRC (United Kingdom) Strategic Equipment
  Grant No. EP/L02263X/1 (EP/M008576/1) and EPSRC (United Kingdom)
  Grant EP/M027791/1.
\end{acknowledgments}


\begin{thebibliography}{52}%
\makeatletter
\providecommand \@ifxundefined [1]{%
 \@ifx{#1\undefined}
}%
\providecommand \@ifnum [1]{%
 \ifnum #1\expandafter \@firstoftwo
 \else \expandafter \@secondoftwo
 \fi
}%
\providecommand \@ifx [1]{%
 \ifx #1\expandafter \@firstoftwo
 \else \expandafter \@secondoftwo
 \fi
}%
\providecommand \natexlab [1]{#1}%
\providecommand \enquote  [1]{``#1''}%
\providecommand \bibnamefont  [1]{#1}%
\providecommand \bibfnamefont [1]{#1}%
\providecommand \citenamefont [1]{#1}%
\providecommand \href@noop [0]{\@secondoftwo}%
\providecommand \href [0]{\begingroup \@sanitize@url \@href}%
\providecommand \@href[1]{\@@startlink{#1}\@@href}%
\providecommand \@@href[1]{\endgroup#1\@@endlink}%
\providecommand \@sanitize@url [0]{\catcode `\\12\catcode `\$12\catcode
  `\&12\catcode `\#12\catcode `\^12\catcode `\_12\catcode `\%12\relax}%
\providecommand \@@startlink[1]{}%
\providecommand \@@endlink[0]{}%
\providecommand \url  [0]{\begingroup\@sanitize@url \@url }%
\providecommand \@url [1]{\endgroup\@href {#1}{\urlprefix }}%
\providecommand \urlprefix  [0]{URL }%
\providecommand \Eprint [0]{\href }%
\providecommand \doibase [0]{http://dx.doi.org/}%
\providecommand \selectlanguage [0]{\@gobble}%
\providecommand \bibinfo  [0]{\@secondoftwo}%
\providecommand \bibfield  [0]{\@secondoftwo}%
\providecommand \translation [1]{[#1]}%
\providecommand \BibitemOpen [0]{}%
\providecommand \bibitemStop [0]{}%
\providecommand \bibitemNoStop [0]{.\EOS\space}%
\providecommand \EOS [0]{\spacefactor3000\relax}%
\providecommand \BibitemShut  [1]{\csname bibitem#1\endcsname}%
\let\auto@bib@innerbib\@empty
\bibitem [{\citenamefont {Sigalas}\ and\ \citenamefont
  {Economou}(1993)}]{sigalas_band_1993}%
  \BibitemOpen
  \bibfield  {author} {\bibinfo {author} {\bibfnamefont {M.}~\bibnamefont
  {Sigalas}}\ and\ \bibinfo {author} {\bibfnamefont {E.~N.}\ \bibnamefont
  {Economou}},\ }\href {\doibase 10.1016/0038-1098(93)90888-T} {\bibfield
  {journal} {\bibinfo  {journal} {Solid State Commun.}\ }\textbf {\bibinfo
  {volume} {86}},\ \bibinfo {pages} {141} (\bibinfo {year} {1993})}\BibitemShut
  {NoStop}%
\bibitem [{\citenamefont {Kushwaha}\ and\ \citenamefont
  {Halevi}(1994)}]{kushwaha_bandgap_1994}%
  \BibitemOpen
  \bibfield  {author} {\bibinfo {author} {\bibfnamefont {M.~S.}\ \bibnamefont
  {Kushwaha}}\ and\ \bibinfo {author} {\bibfnamefont {P.}~\bibnamefont
  {Halevi}},\ }\href {\doibase 10.1063/1.110940} {\bibfield  {journal}
  {\bibinfo  {journal} {Appl. Phys. Lett.}\ }\textbf {\bibinfo {volume} {64}},\
  \bibinfo {pages} {1085} (\bibinfo {year} {1994})}\BibitemShut {NoStop}%
\bibitem [{\citenamefont {Vasseur}\ \emph {et~al.}(2002)\citenamefont
  {Vasseur}, \citenamefont {Deymier}, \citenamefont {Khelif}, \citenamefont
  {Lambin}, \citenamefont {Djafari-Rouhani}, \citenamefont {Akjouj},
  \citenamefont {Dobrzynski}, \citenamefont {Fettouhi},\ and\ \citenamefont
  {Zemmouri}}]{vasseur_phononic_2002}%
  \BibitemOpen
  \bibfield  {author} {\bibinfo {author} {\bibfnamefont {J.~O.}\ \bibnamefont
  {Vasseur}}, \bibinfo {author} {\bibfnamefont {P.~A.}\ \bibnamefont
  {Deymier}}, \bibinfo {author} {\bibfnamefont {A.}~\bibnamefont {Khelif}},
  \bibinfo {author} {\bibfnamefont {P.}~\bibnamefont {Lambin}}, \bibinfo
  {author} {\bibfnamefont {B.}~\bibnamefont {Djafari-Rouhani}}, \bibinfo
  {author} {\bibfnamefont {A.}~\bibnamefont {Akjouj}}, \bibinfo {author}
  {\bibfnamefont {L.}~\bibnamefont {Dobrzynski}}, \bibinfo {author}
  {\bibfnamefont {N.}~\bibnamefont {Fettouhi}}, \ and\ \bibinfo {author}
  {\bibfnamefont {J.}~\bibnamefont {Zemmouri}},\ }\href {\doibase
  10.1103/PhysRevE.65.056608} {\bibfield  {journal} {\bibinfo  {journal} {Phys.
  Rev. E}\ }\textbf {\bibinfo {volume} {65}},\ \bibinfo {pages} {056608} (\bibinfo {year} {2002})}\BibitemShut {NoStop}%
\bibitem [{\citenamefont {Li}\ \emph {et~al.}(2011)\citenamefont {Li},
  \citenamefont {Ni}, \citenamefont {Feng}, \citenamefont {Lu}, \citenamefont
  {He},\ and\ \citenamefont {Chen}}]{li_tunable_2011}%
  \BibitemOpen
  \bibfield  {author} {\bibinfo {author} {\bibfnamefont {X.-F.}\ \bibnamefont
  {Li}}, \bibinfo {author} {\bibfnamefont {X.}~\bibnamefont {Ni}}, \bibinfo
  {author} {\bibfnamefont {L.}~\bibnamefont {Feng}}, \bibinfo {author}
  {\bibfnamefont {M.-H.}\ \bibnamefont {Lu}}, \bibinfo {author} {\bibfnamefont
  {C.}~\bibnamefont {He}}, \ and\ \bibinfo {author} {\bibfnamefont {Y.-F.}\
  \bibnamefont {Chen}},\ }\href {\doibase 10.1103/PhysRevLett.106.084301}
  {\bibfield  {journal} {\bibinfo  {journal} {Phys. Rev. Lett.}\ }\textbf
  {\bibinfo {volume} {106}},\ \bibinfo {pages} {084301} (\bibinfo {year}
  {2011})}\BibitemShut {NoStop}%
\bibitem [{\citenamefont {Chen}\ and\ \citenamefont
  {Chan}(2010)}]{chen_acoustic_2010}%
  \BibitemOpen
  \bibfield  {author} {\bibinfo {author} {\bibfnamefont {H.}~\bibnamefont
  {Chen}}\ and\ \bibinfo {author} {\bibfnamefont {C.~T.}\ \bibnamefont
  {Chan}},\ }\href {\doibase 10.1088/0022-3727/43/11/113001} {\bibfield
  {journal} {\bibinfo  {journal} {J. Phys. D: Appl. Phys.}\ }\textbf {\bibinfo
  {volume} {43}},\ \bibinfo {pages} {113001} (\bibinfo {year}
  {2010})}\BibitemShut {NoStop}%
\bibitem [{\citenamefont {Sukhovich}\ \emph {et~al.}(2008)\citenamefont
  {Sukhovich}, \citenamefont {Jing},\ and\ \citenamefont
  {Page}}]{sukhovich_negative_2008}%
  \BibitemOpen
  \bibfield  {author} {\bibinfo {author} {\bibfnamefont {A.}~\bibnamefont
  {Sukhovich}}, \bibinfo {author} {\bibfnamefont {L.}~\bibnamefont {Jing}}, \
  and\ \bibinfo {author} {\bibfnamefont {J.~H.}\ \bibnamefont {Page}},\ }\href
  {\doibase 10.1103/PhysRevB.77.014301} {\bibfield  {journal} {\bibinfo
  {journal} {Phys. Rev. B}\ }\textbf {\bibinfo {volume} {77}},\ \bibinfo
  {pages} {014301} (\bibinfo {year} {2008})}\BibitemShut {NoStop}%
\bibitem [{\citenamefont {Aspelmeyer}\ \emph {et~al.}(2014)\citenamefont
  {Aspelmeyer}, \citenamefont {Kippenberg},\ and\ \citenamefont
  {Marquardt}}]{aspelmeyer_cavity_2014}%
  \BibitemOpen
  \bibfield  {author} {\bibinfo {author} {\bibfnamefont {M.}~\bibnamefont
  {Aspelmeyer}}, \bibinfo {author} {\bibfnamefont {T.~J.}\ \bibnamefont
  {Kippenberg}}, \ and\ \bibinfo {author} {\bibfnamefont {F.}~\bibnamefont
  {Marquardt}},\ }\href {\doibase 10.1103/RevModPhys.86.1391} {\bibfield
  {journal} {\bibinfo  {journal} {Rev. Mod. Phys.}\ }\textbf {\bibinfo {volume}
  {86}},\ \bibinfo {pages} {1391} (\bibinfo {year} {2014})}\BibitemShut
  {NoStop}%
\bibitem [{\citenamefont {Maldovan}(2013)}]{maldovan_sound_2013}%
  \BibitemOpen
  \bibfield  {author} {\bibinfo {author} {\bibfnamefont {M.}~\bibnamefont
  {Maldovan}},\ }\href {\doibase 10.1038/nature12608} {\bibfield  {journal}
  {\bibinfo  {journal} {Nature}\ }\textbf {\bibinfo {volume} {503}},\ \bibinfo
  {pages} {209} (\bibinfo {year} {2013})}\BibitemShut {NoStop}%
\bibitem [{\citenamefont {Davis}\ and\ \citenamefont
  {Hussein}(2014)}]{davis_nanophononic_2014}%
  \BibitemOpen
  \bibfield  {author} {\bibinfo {author} {\bibfnamefont {B.~L.}\ \bibnamefont
  {Davis}}\ and\ \bibinfo {author} {\bibfnamefont {M.~I.}\ \bibnamefont
  {Hussein}},\ }\href {\doibase 10.1103/PhysRevLett.112.055505} {\bibfield
  {journal} {\bibinfo  {journal} {Phys. Rev. Lett.}\ }\textbf {\bibinfo
  {volume} {112}},\ \bibinfo {pages} {055505} (\bibinfo {year}
  {2014})}\BibitemShut {NoStop}%
\bibitem [{\citenamefont {Torquato}\ and\ \citenamefont
  {Stillinger}(2003)}]{torquato_local_2003}%
  \BibitemOpen
  \bibfield  {author} {\bibinfo {author} {\bibfnamefont {S.}~\bibnamefont
  {Torquato}}\ and\ \bibinfo {author} {\bibfnamefont {F.~H.}\ \bibnamefont
  {Stillinger}},\ }\href {\doibase 10.1103/PhysRevE.68.041113} {\bibfield
  {journal} {\bibinfo  {journal} {Phys. Rev. E}\ }\textbf {\bibinfo {volume}
  {68}},\ \bibinfo {pages} {041113} (\bibinfo {year} {2003})}\BibitemShut
  {NoStop}%
\bibitem [{\citenamefont {Florescu}\ \emph {et~al.}(2009)\citenamefont
  {Florescu}, \citenamefont {Torquato},\ and\ \citenamefont
  {Steinhardt}}]{florescu_designer_2009}%
  \BibitemOpen
  \bibfield  {author} {\bibinfo {author} {\bibfnamefont {M.}~\bibnamefont
  {Florescu}}, \bibinfo {author} {\bibfnamefont {S.}~\bibnamefont {Torquato}},
  \ and\ \bibinfo {author} {\bibfnamefont {P.~J.}\ \bibnamefont {Steinhardt}},\
  }\href {\doibase 10.1073/pnas.0907744106} {\bibfield  {journal} {\bibinfo
  {journal} {PNAS}\ }\textbf {\bibinfo {volume} {106}},\ \bibinfo {pages}
  {20658} (\bibinfo {year} {2009})}\BibitemShut {NoStop}%
\bibitem [{\citenamefont {Man}\ \emph {et~al.}(2013{\natexlab{a}})\citenamefont
  {Man}, \citenamefont {Florescu}, \citenamefont {Williamson}, \citenamefont
  {He}, \citenamefont {Hashemizad}, \citenamefont {Leung}, \citenamefont
  {Liner}, \citenamefont {Torquato}, \citenamefont {Chaikin},\ and\
  \citenamefont {Steinhardt}}]{man_isotropic_2013}%
  \BibitemOpen
  \bibfield  {author} {\bibinfo {author} {\bibfnamefont {W.}~\bibnamefont
  {Man}}, \bibinfo {author} {\bibfnamefont {M.}~\bibnamefont {Florescu}},
  \bibinfo {author} {\bibfnamefont {E.~P.}\ \bibnamefont {Williamson}},
  \bibinfo {author} {\bibfnamefont {Y.}~\bibnamefont {He}}, \bibinfo {author}
  {\bibfnamefont {S.~R.}\ \bibnamefont {Hashemizad}}, \bibinfo {author}
  {\bibfnamefont {B.~Y.~C.}\ \bibnamefont {Leung}}, \bibinfo {author}
  {\bibfnamefont {D.~R.}\ \bibnamefont {Liner}}, \bibinfo {author}
  {\bibfnamefont {S.}~\bibnamefont {Torquato}}, \bibinfo {author}
  {\bibfnamefont {P.~M.}\ \bibnamefont {Chaikin}}, \ and\ \bibinfo {author}
  {\bibfnamefont {P.~J.}\ \bibnamefont {Steinhardt}},\ }\href {\doibase
  10.1073/pnas.1307879110} {\bibfield  {journal} {\bibinfo  {journal} {PNAS}\
  }\textbf {\bibinfo {volume} {110}},\ \bibinfo {pages} {15886} (\bibinfo
  {year} {2013}{\natexlab{a}})}\BibitemShut {NoStop}%
\bibitem [{\citenamefont {Muller}\ \emph {et~al.}(2014)\citenamefont {Muller},
  \citenamefont {Haberko}, \citenamefont {Marichy},\ and\ \citenamefont
  {Scheffold}}]{muller_photonic_2014}%
  \BibitemOpen
  \bibfield  {author} {\bibinfo {author} {\bibfnamefont {N.}~\bibnamefont
  {Muller}}, \bibinfo {author} {\bibfnamefont {J.}~\bibnamefont {Haberko}},
  \bibinfo {author} {\bibfnamefont {C.}~\bibnamefont {Marichy}}, \ and\
  \bibinfo {author} {\bibfnamefont {F.}~\bibnamefont {Scheffold}},\ }\href
  {\doibase 10.1002/adom.201470009} {\bibfield  {journal} {\bibinfo  {journal}
  {Adv. Opt. Mat.}\ }\textbf {\bibinfo {volume} {2}},\ \bibinfo {pages} {104}
  (\bibinfo {year} {2014})}\BibitemShut {NoStop}%
\bibitem [{\citenamefont {Man}\ \emph {et~al.}(2013{\natexlab{b}})\citenamefont
  {Man}, \citenamefont {Florescu}, \citenamefont {Matsuyama}, \citenamefont
  {Yadak}, \citenamefont {Nahal}, \citenamefont {Hashemizad}, \citenamefont
  {Williamson}, \citenamefont {Steinhardt}, \citenamefont {Torquato},\ and\
  \citenamefont {Chaikin}}]{man_photonic_2013}%
  \BibitemOpen
  \bibfield  {author} {\bibinfo {author} {\bibfnamefont {W.}~\bibnamefont
  {Man}}, \bibinfo {author} {\bibfnamefont {M.}~\bibnamefont {Florescu}},
  \bibinfo {author} {\bibfnamefont {K.}~\bibnamefont {Matsuyama}}, \bibinfo
  {author} {\bibfnamefont {P.}~\bibnamefont {Yadak}}, \bibinfo {author}
  {\bibfnamefont {G.}~\bibnamefont {Nahal}}, \bibinfo {author} {\bibfnamefont
  {S.}~\bibnamefont {Hashemizad}}, \bibinfo {author} {\bibfnamefont
  {E.}~\bibnamefont {Williamson}}, \bibinfo {author} {\bibfnamefont
  {P.}~\bibnamefont {Steinhardt}}, \bibinfo {author} {\bibfnamefont
  {S.}~\bibnamefont {Torquato}}, \ and\ \bibinfo {author} {\bibfnamefont
  {P.}~\bibnamefont {Chaikin}},\ }\href {\doibase 10.1364/OE.21.019972}
  {\bibfield  {journal} {\bibinfo  {journal} {Opt. Express}\ }\textbf {\bibinfo
  {volume} {21}},\ \bibinfo {pages} {19972} (\bibinfo {year}
  {2013}{\natexlab{b}})}\BibitemShut {NoStop}%
\bibitem [{\citenamefont {Florescu}\ \emph {et~al.}(2013)\citenamefont
  {Florescu}, \citenamefont {Steinhardt},\ and\ \citenamefont
  {Torquato}}]{florescu_optical_2013}%
  \BibitemOpen
  \bibfield  {author} {\bibinfo {author} {\bibfnamefont {M.}~\bibnamefont
  {Florescu}}, \bibinfo {author} {\bibfnamefont {P.~J.}\ \bibnamefont
  {Steinhardt}}, \ and\ \bibinfo {author} {\bibfnamefont {S.}~\bibnamefont
  {Torquato}},\ }\href {\doibase 10.1103/PhysRevB.87.165116} {\bibfield
  {journal} {\bibinfo  {journal} {Phys. Rev. B}\ }\textbf {\bibinfo {volume}
  {87}},\ \bibinfo {pages} {165116} (\bibinfo {year} {2013})}\BibitemShut
  {NoStop}%
\bibitem [{\citenamefont {Amoah}\ and\ \citenamefont
  {Florescu}(2015{\natexlab{a}})}]{amoah_high-$q$_2015}%
  \BibitemOpen
  \bibfield  {author} {\bibinfo {author} {\bibfnamefont {T.}~\bibnamefont
  {Amoah}}\ and\ \bibinfo {author} {\bibfnamefont {M.}~\bibnamefont
  {Florescu}},\ }\href {\doibase 10.1103/PhysRevB.91.020201} {\bibfield
  {journal} {\bibinfo  {journal} {Phys. Rev. B}\ }\textbf {\bibinfo {volume}
  {91}},\ \bibinfo {pages} {020201} (\bibinfo {year}
  {2015}{\natexlab{a}})}\BibitemShut {NoStop}%
\bibitem [{\citenamefont {Florescu}\ \emph {et~al.}(2010)\citenamefont
  {Florescu}, \citenamefont {Torquato},\ and\ \citenamefont
  {Steinhardt}}]{Florescu2010}%
  \BibitemOpen
  \bibfield  {author} {\bibinfo {author} {\bibfnamefont {M.}~\bibnamefont
  {Florescu}}, \bibinfo {author} {\bibfnamefont {S.}~\bibnamefont {Torquato}},
  \ and\ \bibinfo {author} {\bibfnamefont {P.~J.}\ \bibnamefont {Steinhardt}},\
  }\href {\doibase 10.1063/1.3505322} {\bibfield  {journal} {\bibinfo
  {journal} {Appl. Phys. Lett.}\ }\textbf {\bibinfo {volume} {97}},\ \bibinfo
  {pages} {10} (\bibinfo {year} {2010})},\ \Eprint
  {http://arxiv.org/abs/1011.1698} {arXiv:1011.1698} \BibitemShut {NoStop}%
\bibitem [{\citenamefont {Zhou}\ \emph {et~al.}(2016)\citenamefont {Zhou},
  \citenamefont {Cheng}, \citenamefont {Zhu}, \citenamefont {Sun},\ and\
  \citenamefont {Tsang}}]{zhou_hyperuniform_2016}%
  \BibitemOpen
  \bibfield  {author} {\bibinfo {author} {\bibfnamefont {W.}~\bibnamefont
  {Zhou}}, \bibinfo {author} {\bibfnamefont {Z.}~\bibnamefont {Cheng}},
  \bibinfo {author} {\bibfnamefont {B.}~\bibnamefont {Zhu}}, \bibinfo {author}
  {\bibfnamefont {X.}~\bibnamefont {Sun}}, \ and\ \bibinfo {author}
  {\bibfnamefont {H.~K.}\ \bibnamefont {Tsang}},\ }\href {\doibase
  10.1109/JSTQE.2016.2528125} {\bibfield  {journal} {\bibinfo  {journal} {IEEE
  J. Sel. Top. Quant.}\ }\textbf {\bibinfo {volume} {22}},\ \bibinfo {pages}
  {1} (\bibinfo {year} {2016})}\BibitemShut {NoStop}%
\bibitem [{\citenamefont {Zito}\ \emph {et~al.}(2014)\citenamefont {Zito},
  \citenamefont {Rusciano}, \citenamefont {Pesce}, \citenamefont {Dochshanov},
  \citenamefont {Malafronte}, \citenamefont {Ausanio},\ and\ \citenamefont
  {Sasso}}]{zito_plasmon-enhanced_2014}%
  \BibitemOpen
  \bibfield  {author} {\bibinfo {author} {\bibfnamefont {G.}~\bibnamefont
  {Zito}}, \bibinfo {author} {\bibfnamefont {G.}~\bibnamefont {Rusciano}},
  \bibinfo {author} {\bibfnamefont {G.}~\bibnamefont {Pesce}}, \bibinfo
  {author} {\bibfnamefont {A.}~\bibnamefont {Dochshanov}}, \bibinfo {author}
  {\bibfnamefont {A.}~\bibnamefont {Malafronte}}, \bibinfo {author}
  {\bibfnamefont {G.}~\bibnamefont {Ausanio}}, \ and\ \bibinfo {author}
  {\bibfnamefont {A.}~\bibnamefont {Sasso}},\ }in\ \href {\doibase
  10.1109/MePhoCo.2014.6866481} {\emph {\bibinfo {booktitle} {Photonics
  {Conference}, 2014 {Third} {Mediterranean}}}}\ (\bibinfo {year} {2014})\ pp.\
  \bibinfo {pages} {1--3}\BibitemShut {NoStop}%
\bibitem [{\citenamefont {Degl’Innocenti}\ \emph {et~al.}(2016)\citenamefont
  {Degl’Innocenti}, \citenamefont {Shah}, \citenamefont {Masini},
  \citenamefont {Ronzani}, \citenamefont {Pitanti}, \citenamefont {Ren},
  \citenamefont {Jessop}, \citenamefont {Tredicucci}, \citenamefont {Beere},\
  and\ \citenamefont {Ritchie}}]{deglinnocenti_hyperuniform_2016}%
  \BibitemOpen
  \bibfield  {author} {\bibinfo {author} {\bibfnamefont {R.}~\bibnamefont
  {Degl’Innocenti}}, \bibinfo {author} {\bibfnamefont {Y.~D.}\ \bibnamefont
  {Shah}}, \bibinfo {author} {\bibfnamefont {L.}~\bibnamefont {Masini}},
  \bibinfo {author} {\bibfnamefont {A.}~\bibnamefont {Ronzani}}, \bibinfo
  {author} {\bibfnamefont {A.}~\bibnamefont {Pitanti}}, \bibinfo {author}
  {\bibfnamefont {Y.}~\bibnamefont {Ren}}, \bibinfo {author} {\bibfnamefont
  {D.~S.}\ \bibnamefont {Jessop}}, \bibinfo {author} {\bibfnamefont
  {A.}~\bibnamefont {Tredicucci}}, \bibinfo {author} {\bibfnamefont {H.~E.}\
  \bibnamefont {Beere}}, \ and\ \bibinfo {author} {\bibfnamefont {D.~A.}\
  \bibnamefont {Ritchie}},\ }\href {\doibase 10.1038/srep19325} {\bibfield
  {journal} {\bibinfo  {journal} {Sci. Rep.}\ }\textbf {\bibinfo {volume}
  {6}},\ \bibinfo {pages} {19325} (\bibinfo {year} {2016})}\BibitemShut
  {NoStop}%
\bibitem [{\citenamefont {Zuo-Dong}\ and\ \citenamefont
  {Jian-Chun}(2005)}]{zuo-dong_elastic_2005}%
  \BibitemOpen
  \bibfield  {author} {\bibinfo {author} {\bibfnamefont {Y.}~\bibnamefont
  {Zuo-Dong}}\ and\ \bibinfo {author} {\bibfnamefont {C.}~\bibnamefont
  {Jian-Chun}},\ }\href {\doibase 10.1088/0256-307X/22/4/031} {\bibfield
  {journal} {\bibinfo  {journal} {Chinese Phys. Lett.}\ }\textbf {\bibinfo
  {volume} {22}},\ \bibinfo {pages} {889} (\bibinfo {year} {2005})}\BibitemShut
  {NoStop}%
\bibitem [{\citenamefont {Wagner}\ \emph {et~al.}(2015)\citenamefont {Wagner},
  \citenamefont {Graczykowski}, \citenamefont {Reparaz}, \citenamefont
  {Sachat}, \citenamefont {Sledzinska}, \citenamefont {Alzina},\ and\
  \citenamefont {Torres}}]{wagner_2d_2015}%
  \BibitemOpen
  \bibfield  {author} {\bibinfo {author} {\bibfnamefont {M.~R.}\ \bibnamefont
  {Wagner}}, \bibinfo {author} {\bibfnamefont {B.}~\bibnamefont
  {Graczykowski}}, \bibinfo {author} {\bibfnamefont {J.~S.}\ \bibnamefont
  {Reparaz}}, \bibinfo {author} {\bibfnamefont {A.~E.}\ \bibnamefont {Sachat}},
  \bibinfo {author} {\bibfnamefont {M.}~\bibnamefont {Sledzinska}}, \bibinfo
  {author} {\bibfnamefont {F.}~\bibnamefont {Alzina}}, \ and\ \bibinfo {author}
  {\bibfnamefont {C.~M.~S.}\ \bibnamefont {Torres}},\ }\href
  {http://arxiv.org/abs/1511.07398} {\bibfield  {journal} {\bibinfo  {journal}
  {arXiv:1511.07398 [cond-mat]}\ } (\bibinfo {year} {2015})},\ \bibinfo {note}
  {arXiv: 1511.07398}\BibitemShut {NoStop}%
\bibitem [{\citenamefont {Limonov}\ and\ \citenamefont
  {Rue}(2012)}]{limonov_optical_2012}%
  \BibitemOpen
  \bibfield  {author} {\bibinfo {author} {\bibfnamefont {M.~F.}\ \bibnamefont
  {Limonov}}\ and\ \bibinfo {author} {\bibfnamefont {R.~M. D.~L.}\ \bibnamefont
  {Rue}},\ }\href@noop {} {\emph {\bibinfo {title}
  {Optical {Properties} of {Photonic} {Structures}: {Interplay} of {Order} and
  {Disorder}}}}\ (\bibinfo  {publisher} {CRC Press},\ \bibinfo {year} {2012})\
  \bibinfo {note} {google-Books-ID: 5gABZZ\_KCR0C}\BibitemShut {NoStop}%
\bibitem [{\citenamefont {Sainidou}\ \emph {et~al.}(2005)\citenamefont
  {Sainidou}, \citenamefont {Stefanou},\ and\ \citenamefont
  {Modinos}}]{sainidou_widening_2005}%
  \BibitemOpen
  \bibfield  {author} {\bibinfo {author} {\bibfnamefont {R.}~\bibnamefont
  {Sainidou}}, \bibinfo {author} {\bibfnamefont {N.}~\bibnamefont {Stefanou}},
  \ and\ \bibinfo {author} {\bibfnamefont {A.}~\bibnamefont {Modinos}},\ }\href
  {\doibase 10.1103/PhysRevLett.94.205503} {\bibfield  {journal} {\bibinfo
  {journal} {Phys. Rev. Lett.}\ }\textbf {\bibinfo {volume} {94}},\ \bibinfo
  {pages} {205503} (\bibinfo {year} {2005})}\BibitemShut {NoStop}%
\bibitem [{\citenamefont {Still}\ \emph {et~al.}(2008)\citenamefont {Still},
  \citenamefont {Cheng}, \citenamefont {Retsch}, \citenamefont {Sainidou},
  \citenamefont {Wang}, \citenamefont {Jonas}, \citenamefont {Stefanou},\ and\
  \citenamefont {Fytas}}]{still_simultaneous_2008}%
  \BibitemOpen
  \bibfield  {author} {\bibinfo {author} {\bibfnamefont {T.}~\bibnamefont
  {Still}}, \bibinfo {author} {\bibfnamefont {W.}~\bibnamefont {Cheng}},
  \bibinfo {author} {\bibfnamefont {M.}~\bibnamefont {Retsch}}, \bibinfo
  {author} {\bibfnamefont {R.}~\bibnamefont {Sainidou}}, \bibinfo {author}
  {\bibfnamefont {J.}~\bibnamefont {Wang}}, \bibinfo {author} {\bibfnamefont
  {U.}~\bibnamefont {Jonas}}, \bibinfo {author} {\bibfnamefont
  {N.}~\bibnamefont {Stefanou}}, \ and\ \bibinfo {author} {\bibfnamefont
  {G.}~\bibnamefont {Fytas}},\ }\href {\doibase 10.1103/PhysRevLett.100.194301}
  {\bibfield  {journal} {\bibinfo  {journal} {Phys. Rev. Lett.}\ }\textbf
  {\bibinfo {volume} {100}},\ \bibinfo {pages} {194301} (\bibinfo {year}
  {2008})}\BibitemShut {NoStop}%
\bibitem [{\citenamefont {Chen}\ and\ \citenamefont
  {Wang}(2007)}]{chen_study_2007}%
  \BibitemOpen
  \bibfield  {author} {\bibinfo {author} {\bibfnamefont {A.~L.}\ \bibnamefont
  {Chen}}\ and\ \bibinfo {author} {\bibfnamefont {Y.-S.}\ \bibnamefont
  {Wang}},\ }\href {\doibase 10.1016/j.physb.2006.12.004} {\bibfield  {journal}
  {\bibinfo  {journal} {Physica B: Condensed Matter}\ }\textbf {\bibinfo
  {volume} {392}},\ \bibinfo {pages} {369} (\bibinfo {year}
  {2007})}\BibitemShut {NoStop}%
\bibitem [{\citenamefont {Larose}\ \emph {et~al.}(2004)\citenamefont {Larose},
  \citenamefont {Margerin}, \citenamefont {van Tiggelen},\ and\ \citenamefont
  {Campillo}}]{larose_weak_2004}%
  \BibitemOpen
  \bibfield  {author} {\bibinfo {author} {\bibfnamefont {E.}~\bibnamefont
  {Larose}}, \bibinfo {author} {\bibfnamefont {L.}~\bibnamefont {Margerin}},
  \bibinfo {author} {\bibfnamefont {B.~A.}\ \bibnamefont {van Tiggelen}}, \
  and\ \bibinfo {author} {\bibfnamefont {M.}~\bibnamefont {Campillo}},\ }\href
  {\doibase 10.1103/PhysRevLett.93.048501} {\bibfield  {journal} {\bibinfo
  {journal} {Phys. Rev. Lett.}\ }\textbf {\bibinfo {volume} {93}},\ \bibinfo
  {pages} {048501} (\bibinfo {year} {2004})}\BibitemShut {NoStop}%
\bibitem [{\citenamefont {Shahbazi}\ \emph {et~al.}(2005)\citenamefont
  {Shahbazi}, \citenamefont {Bahraminasab}, \citenamefont {Vaez~Allaei},
  \citenamefont {Sahimi},\ and\ \citenamefont
  {Tabar}}]{shahbazi_localization_2005}%
  \BibitemOpen
  \bibfield  {author} {\bibinfo {author} {\bibfnamefont {F.}~\bibnamefont
  {Shahbazi}}, \bibinfo {author} {\bibfnamefont {A.}~\bibnamefont
  {Bahraminasab}}, \bibinfo {author} {\bibfnamefont {S.~M.}\ \bibnamefont
  {Vaez~Allaei}}, \bibinfo {author} {\bibfnamefont {M.}~\bibnamefont {Sahimi}},
  \ and\ \bibinfo {author} {\bibfnamefont {M.~R.~R.}\ \bibnamefont {Tabar}},\
  }\href {\doibase 10.1103/PhysRevLett.94.165505} {\bibfield  {journal}
  {\bibinfo  {journal} {Phys. Rev. Lett.}\ }\textbf {\bibinfo {volume} {94}},\
  \bibinfo {pages} {165505} (\bibinfo {year} {2005})}\BibitemShut {NoStop}%
\bibitem [{\citenamefont {Tourin}\ \emph {et~al.}(2006)\citenamefont {Tourin},
  \citenamefont {Van Der~Biest},\ and\ \citenamefont
  {Fink}}]{tourin_time_2006-1}%
  \BibitemOpen
  \bibfield  {author} {\bibinfo {author} {\bibfnamefont {A.}~\bibnamefont
  {Tourin}}, \bibinfo {author} {\bibfnamefont {F.}~\bibnamefont {Van
  Der~Biest}}, \ and\ \bibinfo {author} {\bibfnamefont {M.}~\bibnamefont
  {Fink}},\ }\href {\doibase 10.1103/PhysRevLett.96.104301} {\bibfield
  {journal} {\bibinfo  {journal} {Phys. Rev. Lett.}\ }\textbf {\bibinfo
  {volume} {96}},\ \bibinfo {pages} {104301} (\bibinfo {year}
  {2006})}\BibitemShut {NoStop}%
\bibitem [{\citenamefont {Li}\ \emph {et~al.}(2007)\citenamefont {Li},
  \citenamefont {Xu},\ and\ \citenamefont
  {Wang}}]{li_frequency-dependent_2007}%
  \BibitemOpen
  \bibfield  {author} {\bibinfo {author} {\bibfnamefont {F.-M.}\ \bibnamefont
  {Li}}, \bibinfo {author} {\bibfnamefont {M.-Q.}\ \bibnamefont {Xu}}, \ and\
  \bibinfo {author} {\bibfnamefont {Y.-S.}\ \bibnamefont {Wang}},\ }\href
  {\doibase 10.1016/j.ssc.2006.09.019} {\bibfield  {journal} {\bibinfo
  {journal} {Solid State Communications}\ }\textbf {\bibinfo {volume} {141}},\
  \bibinfo {pages} {296} (\bibinfo {year} {2007})}\BibitemShut {NoStop}%
\bibitem [{\citenamefont {Wang}\ \emph {et~al.}(2008)\citenamefont {Wang},
  \citenamefont {Li}, \citenamefont {Huang},\ and\ \citenamefont
  {Wang}}]{wang_propagation_2008}%
  \BibitemOpen
  \bibfield  {author} {\bibinfo {author} {\bibfnamefont {Y.-Z.}\ \bibnamefont
  {Wang}}, \bibinfo {author} {\bibfnamefont {F.-M.}\ \bibnamefont {Li}},
  \bibinfo {author} {\bibfnamefont {W.-H.}\ \bibnamefont {Huang}}, \ and\
  \bibinfo {author} {\bibfnamefont {Y.-S.}\ \bibnamefont {Wang}},\ }\href
  {\doibase 10.1016/j.jmps.2007.07.014} {\bibfield  {journal} {\bibinfo
  {journal} {Journal of the Mechanics and Physics of Solids}\ }\textbf
  {\bibinfo {volume} {56}},\ \bibinfo {pages} {1578} (\bibinfo {year}
  {2008})}\BibitemShut {NoStop}%
\bibitem [{\citenamefont {Lin}\ \emph {et~al.}()\citenamefont {Lin},
  \citenamefont {Knoll},\ and\ \citenamefont {Willey}}]{lin_shape_????}%
  \BibitemOpen
  \bibfield  {author} {\bibinfo {author} {\bibfnamefont {J.}~\bibnamefont
  {Lin}}, \bibinfo {author} {\bibfnamefont {C.}~\bibnamefont {Knoll}}, \ and\
  \bibinfo {author} {\bibfnamefont {C.}~\bibnamefont {Willey}},\ }in\ \href
  {http://arc.aiaa.org/doi/abs/10.2514/6.2006-1896} {\emph {\bibinfo
  {booktitle} {47th {AIAA}/{ASME}/{ASCE}/{AHS}/{ASC} {Structures}, {Structural}
  {Dynamics}, and {Materials} {Conference}}}}\ (\bibinfo  {publisher} {American
  Institute of Aeronautics and Astronautics})\BibitemShut {NoStop}%
\bibitem [{\citenamefont {Davies}\ \emph {et~al.}(2014)\citenamefont {Davies},
  \citenamefont {King}, \citenamefont {Newman}, \citenamefont {Minett},
  \citenamefont {Dunstan},\ and\ \citenamefont
  {Zreiqat}}]{davies_hypothesis:_2014}%
  \BibitemOpen
  \bibfield  {author} {\bibinfo {author} {\bibfnamefont {B.}~\bibnamefont
  {Davies}}, \bibinfo {author} {\bibfnamefont {A.}~\bibnamefont {King}},
  \bibinfo {author} {\bibfnamefont {P.}~\bibnamefont {Newman}}, \bibinfo
  {author} {\bibfnamefont {A.}~\bibnamefont {Minett}}, \bibinfo {author}
  {\bibfnamefont {C.~R.}\ \bibnamefont {Dunstan}}, \ and\ \bibinfo {author}
  {\bibfnamefont {H.}~\bibnamefont {Zreiqat}},\ }\href {\doibase
  10.1038/srep07538} {\bibfield  {journal} {\bibinfo  {journal} {Scientific
  Reports}\ }\textbf {\bibinfo {volume} {4}},\ \bibinfo {pages} {7538}
  (\bibinfo {year} {2014})}\BibitemShut {NoStop}%
\bibitem [{\citenamefont {Shekhawat}\ \emph {et~al.}(2013)\citenamefont
  {Shekhawat}, \citenamefont {Zapperi},\ and\ \citenamefont
  {Sethna}}]{shekhawat_damage_2013}%
  \BibitemOpen
  \bibfield  {author} {\bibinfo {author} {\bibfnamefont {A.}~\bibnamefont
  {Shekhawat}}, \bibinfo {author} {\bibfnamefont {S.}~\bibnamefont {Zapperi}},
  \ and\ \bibinfo {author} {\bibfnamefont {J.~P.}\ \bibnamefont {Sethna}},\
  }\href {\doibase 10.1103/PhysRevLett.110.185505} {\bibfield  {journal}
  {\bibinfo  {journal} {Phys. Rev. Lett.}\ }\textbf {\bibinfo {volume} {110}},\
  \bibinfo {pages} {185505} (\bibinfo {year} {2013})}\BibitemShut {NoStop}%
\bibitem [{\citenamefont {Zen}\ \emph {et~al.}(2014)\citenamefont {Zen},
  \citenamefont {Puurtinen}, \citenamefont {Isotalo}, \citenamefont
  {Chaudhuri},\ and\ \citenamefont {Maasilta}}]{zen_engineering_2014}%
  \BibitemOpen
  \bibfield  {author} {\bibinfo {author} {\bibfnamefont {N.}~\bibnamefont
  {Zen}}, \bibinfo {author} {\bibfnamefont {T.~A.}\ \bibnamefont {Puurtinen}},
  \bibinfo {author} {\bibfnamefont {T.~J.}\ \bibnamefont {Isotalo}}, \bibinfo
  {author} {\bibfnamefont {S.}~\bibnamefont {Chaudhuri}}, \ and\ \bibinfo
  {author} {\bibfnamefont {I.~J.}\ \bibnamefont {Maasilta}},\ }\href {\doibase
  10.1038/ncomms4435} {\bibfield  {journal} {\bibinfo  {journal} {Nat Commun}\
  }\textbf {\bibinfo {volume} {5}},\ \bibinfo {pages} {3435} (\bibinfo {year}
  {2014})}\BibitemShut {NoStop}%
\bibitem [{\citenamefont {O'Connor}\ and\ \citenamefont
  {Lebowitz}(1974)}]{oconnor_heat_1974}%
  \BibitemOpen
  \bibfield  {author} {\bibinfo {author} {\bibfnamefont {A.~J.}\ \bibnamefont
  {O'Connor}}\ and\ \bibinfo {author} {\bibfnamefont {J.~L.}\ \bibnamefont
  {Lebowitz}},\ }\href {\doibase 10.1063/1.1666713} {\bibfield  {journal}
  {\bibinfo  {journal} {Journal of Mathematical Physics}\ }\textbf {\bibinfo
  {volume} {15}},\ \bibinfo {pages} {692} (\bibinfo {year} {1974})}\BibitemShut
  {NoStop}%
\bibitem [{\citenamefont {Maire}\ \emph {et~al.}(2015)\citenamefont {Maire},
  \citenamefont {Anufriev}, \citenamefont {Han}, \citenamefont {Volz},\ and\
  \citenamefont {Nomura}}]{maire_thermal_2015}%
  \BibitemOpen
  \bibfield  {author} {\bibinfo {author} {\bibfnamefont {J.}~\bibnamefont
  {Maire}}, \bibinfo {author} {\bibfnamefont {R.}~\bibnamefont {Anufriev}},
  \bibinfo {author} {\bibfnamefont {H.}~\bibnamefont {Han}}, \bibinfo {author}
  {\bibfnamefont {S.}~\bibnamefont {Volz}}, \ and\ \bibinfo {author}
  {\bibfnamefont {M.}~\bibnamefont {Nomura}},\ }\href
  {http://arxiv.org/abs/1508.04574} {\bibfield  {journal} {\bibinfo  {journal}
  {arXiv:1508.04574 [cond-mat]}\ } (\bibinfo {year} {2015})},\ \bibinfo {note}
  {arXiv: 1508.04574}\BibitemShut {NoStop}%
\bibitem [{\citenamefont {Uche}\ \emph {et~al.}(2004)\citenamefont {Uche},
  \citenamefont {Stillinger},\ and\ \citenamefont
  {Torquato}}]{uche_constraints_2004}%
  \BibitemOpen
  \bibfield  {author} {\bibinfo {author} {\bibfnamefont {O.~U.}\ \bibnamefont
  {Uche}}, \bibinfo {author} {\bibfnamefont {F.~H.}\ \bibnamefont
  {Stillinger}}, \ and\ \bibinfo {author} {\bibfnamefont {S.}~\bibnamefont
  {Torquato}},\ }\href {\doibase 10.1103/PhysRevE.70.046122} {\bibfield
  {journal} {\bibinfo  {journal} {Phys. Rev. E}\ }\textbf {\bibinfo {volume}
  {70}},\ \bibinfo {pages} {046122} (\bibinfo {year} {2004})}\BibitemShut
  {NoStop}%
\bibitem [{\citenamefont {Chen}\ \emph {et~al.}(2010)\citenamefont {Chen},
  \citenamefont {Wang}, \citenamefont {Li},\ and\ \citenamefont
  {Zhang}}]{chen_localisation_2010}%
  \BibitemOpen
  \bibfield  {author} {\bibinfo {author} {\bibfnamefont {A.-L.}\ \bibnamefont
  {Chen}}, \bibinfo {author} {\bibfnamefont {Y.-S.}\ \bibnamefont {Wang}},
  \bibinfo {author} {\bibfnamefont {J.-B.}\ \bibnamefont {Li}}, \ and\ \bibinfo
  {author} {\bibfnamefont {C.}~\bibnamefont {Zhang}},\ }\href {\doibase
  10.1080/17455030903394568} {\bibfield  {journal} {\bibinfo  {journal} {Waves
  in Random and Complex Media}\ }\textbf {\bibinfo {volume} {20}},\ \bibinfo
  {pages} {104} (\bibinfo {year} {2010})}\BibitemShut {NoStop}%
\bibitem [{\citenamefont {Batten}\ \emph {et~al.}(2008)\citenamefont {Batten},
  \citenamefont {Stillinger},\ and\ \citenamefont
  {Torquato}}]{batten_classical_2008}%
  \BibitemOpen
  \bibfield  {author} {\bibinfo {author} {\bibfnamefont {R.~D.}\ \bibnamefont
  {Batten}}, \bibinfo {author} {\bibfnamefont {F.~H.}\ \bibnamefont
  {Stillinger}}, \ and\ \bibinfo {author} {\bibfnamefont {S.}~\bibnamefont
  {Torquato}},\ }\href {\doibase 10.1063/1.2961314} {\bibfield  {journal}
  {\bibinfo  {journal} {J. Appl. Phys.}\ }\textbf {\bibinfo {volume} {104}},\
  \bibinfo {pages} {033504} (\bibinfo {year} {2008})}\BibitemShut {NoStop}%
\bibitem [{\citenamefont {White}(1958)}]{white_elastic_1958}%
  \BibitemOpen
  \bibfield  {author} {\bibinfo {author} {\bibfnamefont {R.~M.}\ \bibnamefont
  {White}},\ }\href {\doibase 10.1121/1.1909759} {\bibfield  {journal}
  {\bibinfo  {journal} {The Journal of the Acoustical Society of America}\
  }\textbf {\bibinfo {volume} {30}},\ \bibinfo {pages} {771} (\bibinfo {year}
  {1958})}\BibitemShut {NoStop}%
\bibitem [{\citenamefont {Malvern}(1969)}]{malvern_introduction_1969}%
  \BibitemOpen
  \bibfield  {author} {\bibinfo {author} {\bibfnamefont {L.~E.}\ \bibnamefont
  {Malvern}},\ }\href@noop {} {\emph {\bibinfo {title} {Introduction to the
  Mechanics of a Continuous Medium}}}\ (\bibinfo  {publisher} {Englewood
  Cliffs, N.J.: Prentice-Hall},\ \bibinfo {year} {1969})\BibitemShut {NoStop}%
\bibitem [{\citenamefont {Sainidou}\ \emph {et~al.}(2004)\citenamefont
  {Sainidou}, \citenamefont {Stefanou},\ and\ \citenamefont
  {Modinos}}]{sainidou_greens_2004}%
  \BibitemOpen
  \bibfield  {author} {\bibinfo {author} {\bibfnamefont {R.}~\bibnamefont
  {Sainidou}}, \bibinfo {author} {\bibfnamefont {N.}~\bibnamefont {Stefanou}},
  \ and\ \bibinfo {author} {\bibfnamefont {A.}~\bibnamefont {Modinos}},\ }\href
  {\doibase 10.1103/PhysRevB.69.064301} {\bibfield  {journal} {\bibinfo
  {journal} {Phys. Rev. B}\ }\textbf {\bibinfo {volume} {69}},\ \bibinfo
  {pages} {064301} (\bibinfo {year} {2004})}\BibitemShut {NoStop}%
\bibitem [{\citenamefont {Kafesaki}\ \emph {et~al.}(1995)\citenamefont
  {Kafesaki}, \citenamefont {Sigalas},\ and\ \citenamefont
  {Economou}}]{kafesaki_elastic_1995}%
  \BibitemOpen
  \bibfield  {author} {\bibinfo {author} {\bibfnamefont {M.}~\bibnamefont
  {Kafesaki}}, \bibinfo {author} {\bibfnamefont {M.~M.}\ \bibnamefont
  {Sigalas}}, \ and\ \bibinfo {author} {\bibfnamefont {E.~N.}\ \bibnamefont
  {Economou}},\ }\href {\doibase 10.1016/0038-1098(95)00444-0} {\bibfield
  {journal} {\bibinfo  {journal} {Solid State Commun.}\ }\textbf {\bibinfo
  {volume} {96}},\ \bibinfo {pages} {285} (\bibinfo {year} {1995})}\BibitemShut
  {NoStop}%
\bibitem [{\citenamefont {Tsitrin}\ \emph {et~al.}(2015)\citenamefont
  {Tsitrin}, \citenamefont {Williamson}, \citenamefont {Amoah}, \citenamefont
  {Nahal}, \citenamefont {Chan}, \citenamefont {Florescu},\ and\ \citenamefont
  {Man}}]{tsitrin_unfolding_2015}%
  \BibitemOpen
  \bibfield  {author} {\bibinfo {author} {\bibfnamefont {S.}~\bibnamefont
  {Tsitrin}}, \bibinfo {author} {\bibfnamefont {E.~P.}\ \bibnamefont
  {Williamson}}, \bibinfo {author} {\bibfnamefont {T.}~\bibnamefont {Amoah}},
  \bibinfo {author} {\bibfnamefont {G.}~\bibnamefont {Nahal}}, \bibinfo
  {author} {\bibfnamefont {H.~L.}\ \bibnamefont {Chan}}, \bibinfo {author}
  {\bibfnamefont {M.}~\bibnamefont {Florescu}}, \ and\ \bibinfo {author}
  {\bibfnamefont {W.}~\bibnamefont {Man}},\ }\href {\doibase 10.1038/srep13301}
  {\bibfield  {journal} {\bibinfo  {journal} {Sci. Rep.}\ }\textbf {\bibinfo
  {volume} {5}},\ \bibinfo {pages} {13301} (\bibinfo {year}
  {2015})}\BibitemShut {NoStop}%
\bibitem{_realizing_2010}
  \BibitemOpen
  \bibfield {author} {\bibinfo {author} {\bibfnamefont {D.}~\bibnamefont {Goettler}}\ ,
    \ \bibinfo {author} {\bibfnamefont {M.}~\bibnamefont {Su}},
    \ \bibinfo {author} {\bibfnamefont {Z.}~\bibnamefont {Leseman}},
    \ \bibinfo {author} {\bibfnamefont {Y.}~\bibnamefont {Soliman}},
    \ \bibinfo {author} {\bibfnamefont {R.}~\bibnamefont {Olsson}},
    \ \bibinfo {author} {\bibfnamefont {I.}~\bibnamefont {El-Kady}},\ }
  \href {\doibase 10.1063/1.3475987} {\bibfield
    {journal} {\bibinfo  {journal} {J. Appl. Phys.}\ }\textbf {\bibinfo
      {volume} {108}},\ \bibinfo {pages} {084505} (\bibinfo {year}
    {2010})}\BibitemShut {NoStop}%
\bibitem [{\citenamefont {Khelif}\ \emph {et~al.}(2004)\citenamefont {Khelif},
  \citenamefont {Choujaa}, \citenamefont {Benchabane}, \citenamefont
  {Djafari-Rouhani},\ and\ \citenamefont {Laude}}]{khelif_guiding_2004}%
  \BibitemOpen
  \bibfield  {author} {\bibinfo {author} {\bibfnamefont {A.}~\bibnamefont
  {Khelif}}, \bibinfo {author} {\bibfnamefont {A.}~\bibnamefont {Choujaa}},
  \bibinfo {author} {\bibfnamefont {S.}~\bibnamefont {Benchabane}}, \bibinfo
  {author} {\bibfnamefont {B.}~\bibnamefont {Djafari-Rouhani}}, \ and\ \bibinfo
  {author} {\bibfnamefont {V.}~\bibnamefont {Laude}},\ }\href {\doibase
  10.1063/1.1757642} {\bibfield  {journal} {\bibinfo  {journal} {Appl. Phys.
  Lett.}\ }\textbf {\bibinfo {volume} {84}},\ \bibinfo {pages} {4400} (\bibinfo
  {year} {2004})}\BibitemShut {NoStop}%
\bibitem [{\citenamefont {Khelif}\ \emph {et~al.}(2003)\citenamefont {Khelif},
  \citenamefont {Choujaa}, \citenamefont {Djafari-Rouhani}, \citenamefont
  {Wilm}, \citenamefont {Ballandras},\ and\ \citenamefont
  {Laude}}]{khelif_trapping_2003}%
  \BibitemOpen
  \bibfield  {author} {\bibinfo {author} {\bibfnamefont {A.}~\bibnamefont
  {Khelif}}, \bibinfo {author} {\bibfnamefont {A.}~\bibnamefont {Choujaa}},
  \bibinfo {author} {\bibfnamefont {B.}~\bibnamefont {Djafari-Rouhani}},
  \bibinfo {author} {\bibfnamefont {M.}~\bibnamefont {Wilm}}, \bibinfo {author}
  {\bibfnamefont {S.}~\bibnamefont {Ballandras}}, \ and\ \bibinfo {author}
  {\bibfnamefont {V.}~\bibnamefont {Laude}},\ }\href {\doibase
  10.1103/PhysRevB.68.214301} {\bibfield  {journal} {\bibinfo  {journal} {Phys.
  Rev. B}\ }\textbf {\bibinfo {volume} {68}},\ \bibinfo {pages} {214301}
  (\bibinfo {year} {2003})}\BibitemShut {NoStop}%
\bibitem [{\citenamefont {Amoah}\ and\ \citenamefont
  {Florescu}(2015{\natexlab{b}})}]{amoah_hyperuniform_2015}%
  \BibitemOpen
  \bibfield  {author} {\bibinfo {author} {\bibfnamefont {T.}~\bibnamefont
  {Amoah}}\ and\ \bibinfo {author} {\bibfnamefont {M.}~\bibnamefont
  {Florescu}}\ }(\bibinfo  {publisher} {Proc. SPIE},\ \bibinfo {year} {2015})\
  p.\ \bibinfo {pages} {95460F}\BibitemShut {NoStop}%
\bibitem [{\citenamefont {Eichenfield}\ \emph {et~al.}(2009)\citenamefont
  {Eichenfield}, \citenamefont {Chan}, \citenamefont {Camacho}, \citenamefont
  {Vahala},\ and\ \citenamefont {Painter}}]{eichenfield_optomechanical_2009}%
  \BibitemOpen
  \bibfield  {author} {\bibinfo {author} {\bibfnamefont {M.}~\bibnamefont
  {Eichenfield}}, \bibinfo {author} {\bibfnamefont {J.}~\bibnamefont {Chan}},
  \bibinfo {author} {\bibfnamefont {R.~M.}\ \bibnamefont {Camacho}}, \bibinfo
  {author} {\bibfnamefont {K.~J.}\ \bibnamefont {Vahala}}, \ and\ \bibinfo
  {author} {\bibfnamefont {O.}~\bibnamefont {Painter}},\ }\href {\doibase
  10.1038/nature08524} {\bibfield  {journal} {\bibinfo  {journal} {Nature}\
  }\textbf {\bibinfo {volume} {462}},\ \bibinfo {pages} {78} (\bibinfo {year}
  {2009})}\BibitemShut {NoStop}%
\bibitem [{\citenamefont {Lucklum}\ \emph {et~al.}(2013)\citenamefont
  {Lucklum}, \citenamefont {Zubtsov},\ and\ \citenamefont
  {Oseev}}]{lucklum_phoxonic_2013}%
  \BibitemOpen
  \bibfield  {author} {\bibinfo {author} {\bibfnamefont {R.}~\bibnamefont
  {Lucklum}}, \bibinfo {author} {\bibfnamefont {M.}~\bibnamefont {Zubtsov}}, \
  and\ \bibinfo {author} {\bibfnamefont {A.}~\bibnamefont {Oseev}},\ }\href
  {\doibase 10.1007/s00216-013-7093-9} {\bibfield  {journal} {\bibinfo
  {journal} {Anal Bioanal Chem}\ }\textbf {\bibinfo {volume} {405}},\ \bibinfo
  {pages} {6497} (\bibinfo {year} {2013})}\BibitemShut {NoStop}%
\bibitem [{\citenamefont {Djafari-Rouhani}\ \emph {et~al.}(2016)\citenamefont
  {Djafari-Rouhani}, \citenamefont {El-Jallal},\ and\ \citenamefont
  {Pennec}}]{djafari-rouhani_phoxonic_2016}%
  \BibitemOpen
  \bibfield  {author} {\bibinfo {author} {\bibfnamefont {B.}~\bibnamefont
  {Djafari-Rouhani}}, \bibinfo {author} {\bibfnamefont {S.}~\bibnamefont
  {El-Jallal}}, \ and\ \bibinfo {author} {\bibfnamefont {Y.}~\bibnamefont
  {Pennec}},\ }\href {\doibase 10.1016/j.crhy.2016.02.001} {\bibfield
  {journal} {\bibinfo  {journal} {Comptes Rendus Physique}\ }\bibinfo {series}
  {Phononic crystals / {Cristaux} phononiques},\ \textbf {\bibinfo {volume}
  {17}},\ \bibinfo {pages} {555} (\bibinfo {year} {2016})}\BibitemShut
  {NoStop}%
\bibitem [{\citenamefont {III}\ and\ \citenamefont
  {El-Kady}(2009)}]{iii_microfabricated_2009}%
  \BibitemOpen
  \bibfield  {author} {\bibinfo {author} {\bibfnamefont {R.~H.~O.}\
  \bibnamefont {III}}\ and\ \bibinfo {author} {\bibfnamefont {I.}~\bibnamefont
  {El-Kady}},\ }\href {\doibase 10.1088/0957-0233/20/1/012002} {\bibfield
  {journal} {\bibinfo  {journal} {Meas. Sci. Technol.}\ }\textbf {\bibinfo
  {volume} {20}},\ \bibinfo {pages} {012002} (\bibinfo {year}
  {2009})}\BibitemShut {NoStop}%
\end{thebibliography}
\end{document}